\documentclass{statsoc}
\usepackage{amssymb}
\usepackage{graphicx}
\usepackage{amsmath}
\usepackage{latexsym}
\RequirePackage{float,epsfig,multirow,rotating,times}

\usepackage{multirow}
\usepackage{float}

\floatstyle{plaintop}
\restylefloat{table}
\usepackage{color}

\usepackage{hyperref}
\usepackage{listings}
\usepackage{xcolor}

\lstdefinestyle{mystyle}{
    backgroundcolor = \color{lightgray},
    frame=single,
    framerule=0pt,
    framesep=10pt,
    basicstyle=\ttfamily\footnotesize,
    breakatwhitespace=false,         
    breaklines=true,                 
    captionpos=b,                    
    keepspaces=true,                 
    numbers=left,                    
    numbersep=5pt,                  
    showspaces=false,                
    showstringspaces=false,
    showtabs=false,                  
    tabsize=2
}

\newcommand{\boldnu}{\mbox{\boldmath $\nu$}}
\newcommand{\boldtheta}{\mbox{\boldmath $\theta$}}
\newcommand{\boldeta}{\mbox{\boldmath $\eta$}}
\newcommand{\boldbeta}{\mbox{\boldmath $\beta$}}
\newcommand{\boldalpha}{\mbox{\boldmath $\alpha$}}
\newcommand{\boldmu}{\mbox{\boldmath $\mu$}}
\newcommand{\boldgamma}{\mbox{\boldmath $\gamma$}}
\newcommand{\bolddelta}{\mbox{\boldmath $\delta$}}
\newcommand{\boldphi}{\mbox{\boldmath $\phi$}}
\newcommand{\boldpsi}{\mbox{\boldmath $\psi$}}
\newcommand{\boldomega}{\mbox{\boldmath $\omega$}}

\newcommand{\boldy}{{\boldsymbol y}}

\newcommand{\boldzero}{\boldsymbol{0}}

\title[Bayesian hierarchical modelling approaches for immunization coverage estimation] {Bayesian hierarchical modelling approaches for combining information from multiple data sources to produce annual estimates of national immunization coverage}

\author[Utazi \it{et al.}]
       {C. Edson Utazi}
       \coaddress{WorldPop, School of Geography and Environmental Science, University of Southampton, UK}
       \email{C.E.Utazi@soton.ac.uk}
 \author[Utazi \it{et al.}]
       {Warren C. Jochem}
       \address{WorldPop, School of Geography and Environmental Science, University of Southampton, UK}
   \author[Utazi \it{et al.}]
       {Marta Gacic-Dobo}
       \address{World Health Organization, Geneva, Switzerland} 
   \author[Utazi \it{et al.}]
       {Padraic Murphy}
       \address{United Nations Children's Fund, New York, USA}
    \author[Utazi \it{et al.}]
       {Sujit K. Sahu}
   \address{Mathematical Sciences, University of Southampton, UK}
    \author[Utazi \it{et al.}]
       {M. Carolina Danovaro-Holliday}
       \address{World Health Organization, Geneva, Switzerland} 
    \author[Utazi \it{et al.}]
       {Andrew J. Tatem}
       \address{WorldPop, School of Geography and Environmental Science, University of Southampton, United Kingdom}  
    
\begin{document}

\begin{abstract}
Estimates of national immunization coverage are crucial for guiding policy and decision-making in national immunization programs and setting the global immunization agenda. WHO and UNICEF estimates of national immunization coverage (WUENIC) are produced annually for various vaccine-dose combinations and all WHO Member States using information from multiple data sources and a deterministic computational logic approach. This approach, however, is incapable of characterizing the uncertainties inherent in coverage measurement and estimation. It also provides no statistically principled way of exploiting and accounting for the interdependence in immunization coverage data collected for multiple vaccines, countries and time points. Here, we develop Bayesian hierarchical modeling approaches for producing accurate estimates of national immunization coverage and their associated uncertainties. We propose and explore two candidate models: a balanced data single likelihood (BDSL) model and an irregular data multiple likelihood (IDML) model, both of which differ in their handling of missing data and characterization of the uncertainties associated with the multiple input data sources. We provide a simulation study that demonstrates a high degree of accuracy of the estimates produced by the proposed models, and which also shows that the IDML model is the better model. We apply the methodology to produce coverage estimates for select vaccine-dose combinations for the period 2000-2019. A contributed R package {\tt imcover} implementing the No-U-Turn Sampler (NUTS) in the Stan  programming language enhances the utility and reproducibility of the methodology. 


\end{abstract}

\section{Introduction}
\label{sec:intro}
Accurate estimates of immunization coverage are required at the global, regional, national and subnational levels to inform policies and guide interventions aimed at improving coverage levels and accelerating progress towards disease elimination and eradication \citep{Burton2009,IA2030,GAVI5.0}. In particular, globally comparable estimates of national immunization coverage (ENIC) are crucial for profiling countries and understanding where attention should be focused to strengthen global immunization service delivery. Immunization coverage is also an important indicator for measuring and monitoring progress towards the goals and targets set out in global policy frameworks such as the Sustainable Development Goals (SDGs) \citep{SDG2015} and the Immunization Agenda 2030 (IA2030) \citep{IA2030}. 
Since 1998, the World Health Organization (WHO) and the United Nations International Children's Emergency Fund (UNICEF) have jointly published estimates of national immunization coverage annually for 195 countries and territories and different vaccine-dose combinations \citep{Burton2012,Burton2009,Danovaro-Hollidayetal2021}. 

WHO and UNICEF estimates of national immunization coverage (WUENIC) are produced through 
integrating information from multiple data sources using a computational logic approach \citep{Burton2009,Burton2012}. Fundamentally, the approach uses deterministic ad hoc estimation rules to supplement administrative and official ENIC reported to WHO and UNICEF \citep{WHOUNICEF2022} with estimates obtained via nationally representative 
household surveys. The approach is implemented on a country-by-country basis, enabling the incorporation of expert knowledge and country-specific adjustments (e.g., vaccine stockouts, conflict and other data quality assessments). Although the approach encourages reproducibility, representing an improvement over informal, manual estimation procedures, it is incapable of characterizing the uncertainties inherent in the multiple input data sets and those associated with the output coverage estimates. Also, the approach provides no statistically principled way of accounting for and exploiting the different sources of variation or dependence that may exist in immunization coverage data collected for multiple vaccines, countries and years. The entire time series produced by the approach is usually updated when new data become available; however, the approach is also unsuited for producing estimates of immunization coverage for future time points which may be useful to guide program planning. To address these challenges, WHO and UNICEF made a call in 2019 for alternative approaches for producing ENIC. 

Bayesian hierarchical modeling (BHM) approaches \citep{cressieandwikle,Sahubook} offer a robust, flexible framework to combine information from multiple data sources to estimate a phenomenon of interest, whilst accounting for the full range of uncertainty present in these data sources. BHM approaches have been widely applied in different thematic areas ranging from environmental studies to health and medicine  \citep[see, e.g.,][]{Sahubook}. In immunization coverage estimation, the use of BHM approaches has mostly focused on producing estimates of coverage at a high resolution (e.g, 1 km or 5 km grid squares covering an area of interest) and for subnational administrative areas (e.g. the district level), where the intermediate levels of the model relate to characterizing the spatial and spatiotemporal variation in the data using geostatistical and conditional autoregressive models \citep{LBD2021, Utazi2020pccs, Utazisim2021}. Similar applications also exist in the estimation of other health and development indicators \citep[e.g.,][]{sahu2006,Burstein2019,Giorgietal2021}. Whilst estimates of immunization coverage produced using these approaches are extremely valuable for uncovering the spatial heterogeneities in coverage that often exist within countries, estimates of national immunization coverage and other indicators still remain the benchmark for policy and decision making, resource allocation and monitoring and evaluating progress at the global level.

Here, we develop novel BHM approaches for producing ENIC as an alternative or a complement to the WUENIC computational logic approach. Similar to WUENIC approach, our methodology utilizes multiple input data and enables the incorporation of expert opinions and judgments which can be implemented before or during model-fitting. However, it also crucially provides a mechanism to leverage various sources of dependence in the input data to improve the estimation of coverage and associated uncertainties. We propose and explore two candidate models, namely a balanced data single likelihood (BDSL) model and an irregular data multiple likelihood (IDML) model, both of which differ in their 
handling of missing data and
characterization of the uncertainties arising from the multiple input data sources. The proposed models are implemented for each of the six WHO regions, but can also be implemented at the country level, although regional models provide a richer framework to exploit inter-country variation. The methodology is supported by the development and Github publication of a contributed R package {\tt imcover} to enhance its utility and to encourage reproducibility.

In what follows, we present and explore the data used in this work. We also outline the steps taken to process the data, including a recall-bias adjustment for survey data as in the WUENIC approach \citep{Burton2009,Burton2012,WHORB2015}. We then proceed to present the proposed methodology and its implementation in a Bayesian framework using the Stan package in R \citep{rstan}. A simulation study assessing both in-sample and out-of-sample predictive performance of the methodology is presented in Section \ref{sec:simstudy}. Modelled estimates of coverage and associated uncertainties are presented and discussed in Section \ref{sec:application}, as well as comparisons with corresponding WUENIC estimates (2020 revision published in 2021). In Section \ref{sec:software}, a description of the accompanying {\tt imcover} software package is provided. We conclude with a discussion on both the methodology and the modelled outputs and outline directions for future work.

\section{Data description and processing}
\label{sec:data}

We assembled publicly available, aggregate data on national immunization coverage from the WHO Immunization Data Portal (\url{https://immunizationdata.who.int}; accessed on March 2, 2022). The data portal provides access to information from the WHO/UNICEF Joint Reporting Form on Immunization (JRF) \citep{WHOUNICEF2022}. We collect data from three main sources of information on national-level immunization coverage:
\begin{itemize}
\item[(i)] reported administrative coverage data (admin);
\item[(ii)] country-reported official coverage estimates (official);
\item[(iii)] household surveys of vaccination coverage (survey).
\end{itemize}
Administrative coverage data are estimates of vaccination coverage which countries report annually to WHO and UNICEF through the JRF. Official estimates are coverage reports which have been independently reviewed by countries against other datasets and represent their assessment of the most likely coverage. In most cases, official estimates are the same as administrative estimates, but in some cases, are based on surveys or ``corrections'' for known inaccuracies in the admin estimates, e.g., incomplete reporting. The survey data comes from nationally representative household survey data used by WUENIC. There are three main survey sources: the Expanded Programme on Immunization (EPI) cluster survey \citep{WHOsurv2018} and surveys using previous WHO recommendations for vaccination coverage surveys \citep{DANOVAROHOLLIDAY20185150}; the Demographic and Health Surveys \citep{ICFInt}; and the Multiple Indicator Cluster Survey (MICS) \citep{MICS}. For each of these sources (admin, official, and survey), we obtain annual data on five vaccines for the period 2000 - 2019: diphtheria-tetanus-pertussis-containing vaccine doses 1 and 3 (DTP1 and DTP3), measles containing vaccine doses 1 and 2 (MCV1 and MCV2), and pneumococcal conjugate vaccine dose 3 (PCV3).

In addition to the coverage information, we obtain mid-year estimates of countries' population from the UN Population Division 2019 revision \citep{WPP2019}. These population data serve as denominators to estimate the percentage of the population covered by administered vaccine doses. Therefore, the age cohorts of the population data correspond to the target population to receive the selected vaccines. In the case of DTP1, DTP3, MCV1, and PCV3, this is surviving infants (i.e., under 1 year old, even when MCV1 is recommended in the second year of life in some countries), while the target age for MCV2 depends on the national immunization schedule \citep{WHOIS}. Finally, we collect information on the year of vaccine introduction (yovi) from the WHO repository. Since MCV2 and PCV3 vaccines were not used across all countries in this period, we restrict our analyses to periods when vaccines were fully rolled-out.

All WHO Member States spread across six WHO regions (namely, African Region (AFR), Region of the Americas (AMR), Eastern Mediterranean Region (EMR), European region (EUR), South-East Asian Region (SEAR) and Western Pacific Region (WPR)) and for which data were available are included in this work (see supplementary Figure 2). After obtaining the raw datasets described above, a multi-step data cleaning and harmonisation process was implemented to create analysis-ready data for model fitting. In the first stage of processing, standard vaccine-source-specific input data files were created for the study period from the raw data by extracting the relevant data required for the analysis. This was followed by a recall-bias adjustment step implemented for DTP3 and PCV3 using survey data \citep{WHORB2015}. This is detailed in the supplementary file. Further, to ensure that modelled estimates of DTP1 and DTP3 were consistent, i.e. that DTP1 is greater than or equal to DTP3 for all country-vaccine-time combinations, we opted to model DTP1 and the ratio of DTP1 and DTP3, i.e. DTP3/DTP1 (later on in the modelling step, we converted these ratios to DTP3 estimates by multiplying corresponding DTP1 and DTP3/DTP1 estimates). To calculate the ratios, we first preserved the differential between DTP1 and DTP3 where estimates of the former were greater than 100 by calculating the ratios using the original input data. We then rounded down those estimates greater than 100 to 99.9\% (this is necessary for the logit-transformation step in Section~\ref{sec:method}) and adjusted corresponding DTP3 estimates using the ratios obtained previously. Finally, for each country, vaccine, time and data source, we scaled the coverage data to the unit interval and logit-transformed these as a final pre-modelling step. During this step, estimates of other vaccines greater than 1 were set to 0.999 and, to avoid undefined values in the transformation, any estimates equal to zero were adjusted to 0.001. 
The results of these processing steps is a collection of country-specific, immunization coverage estimates from three major sources of information. The three sources of information (admin, official and survey) form three time series, although with different levels of variation over time and observation time points and completeness. The challenge of the proposed modelling approach outlined in Section \ref{sec:method} is to draw information from these different sources to estimate a true, but unobserved, national immunisation coverage estimate. 


\begin{table}[ht]
\caption{Summary of processed national immunization coverage data for the period 2000 – 2019 for all WHO Member States and data sources. Shown in the second column are numbers of countries with input data and numbers of non-missing data points. \vspace{0.2cm}}
\centering
\begin{tabular}{c r r r r r r r r} \hline \hline \hline
 {\bf Vaccine}\slash & {\bf No. of $\quad$} & \multicolumn{7}{c}{\textbf {Summary statistics (\%)}}\\[0.5 ex]
 \cline{3-9}
  {\bf data} & {\bf countries}/ & {\bf Min.} & {\bf Q1} & {\bf Med.}   & {\bf Mean}   & {\bf Std.}  & {\bf Q3}	& {\bf Max.}\\
  {\bf source} & {\bf data points}  &   &       &       &       &{\bf dev.}  &      & \\
\hline
DTP1	&189 (6787)	 &10.00	&89.00	&96.00	&91.43	&11.57	&99.00	&99.99\\ 
DTP3	&189 (6763)	 &0.92	&81.00	&91.00	&85.75	&15.16	&96.00	&99.98\\
MCV1	&194 (7577)	 &2.10	&81.00	&92.00	&86.47	&14.88	&97.00	&99.99\\
MCV2	&176 (4191)	 &1.00	&78.00	&91.52	&84.03	&18.70	&96.97	&99.99\\
PCV3	&148 (2144)	 &1.00	&76.00	&89.00	&79.97	&23.26	&95.00	&99.99\\
Admin	&186 (12567) &0.92	&83.36	&93.00	&87.21	&15.79	&97.40	&99.99\\
Official&194 (12872) &1.00	&84.00	&93.00	&87.44	 &15.27	&97.20	&99.99\\
Survey	&131 (2023)	 &2.10	&67.95	&83.90	&78.03	&19.35	&93.00	&99.90\\
All vaccines/ &      &	    &	    &	    &	    &	    &   	&\\
sources &194 (27462)	&0.92	&82.20	&92.53	&86.64	&16.02	&97.00	&99.99\\ \hline \hline \hline
\end{tabular}
\label{tab:datasummary}
\end{table}

\begin{figure}[htb]
    \centering
 \includegraphics[width=1.1 \textwidth]{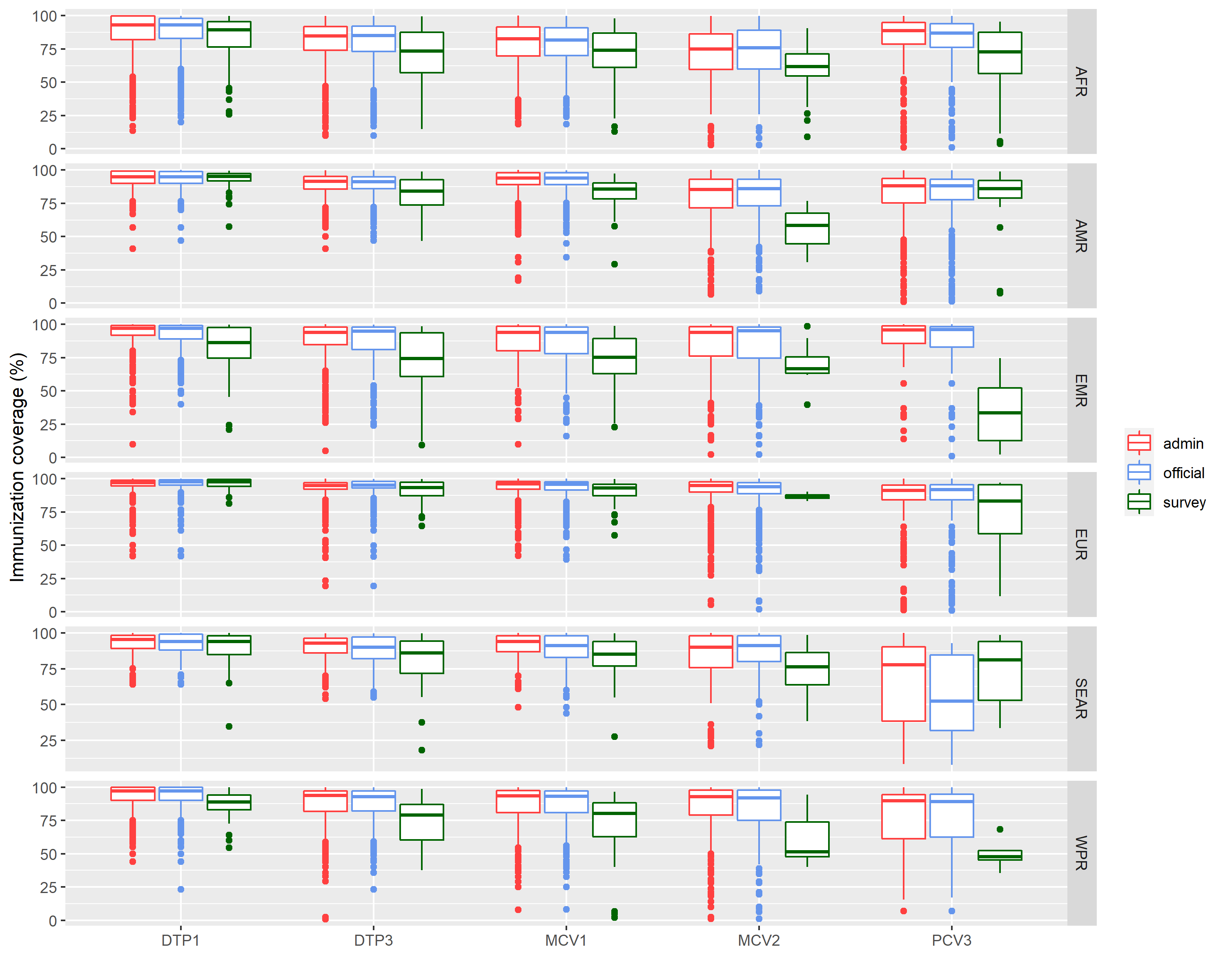} 
 \caption{Distribution of processed national immunization coverage data by WHO region and data source.
\label{fig:regboxplot}} 
\end{figure}

Summaries of the processed data are presented in Table~\ref{tab:datasummary} at the global level and Figure \ref{fig:regboxplot} for each WHO region. For illustrations of the patterns of missingness in these data, see supplementary Figure 3 and Figure \ref{fig:multilikfig}. In all, about 46\% of the data were from administrative sources, while 47\% and 7\% were from official sources and surveys, respectively. 186 countries had administrative data, 194 countries had official data, while 131 had survey data for at least one of the five vaccines at any time point during the study period. When considering data from all sources at the global level, Table \ref{tab:datasummary} shows that PCV3 had the smallest number of data points, the most variability and the lowest coverage due to being only universally recommended by WHO in 2009. DTP1 had the highest coverage and lowest variation, whereas DTP3 and MCV1 had very similar distributions.

Further, Figure \ref{fig:regboxplot} shows that within each region, coverage data obtained from surveys were more likely to be lower on average than data obtained from official and administrative sources (except in the case of PCV3 in SEAR). We also observe that administrative and official data have very similar distributions as expected. We note that the processing steps outlined here have been developed in conjunction with the WHO/UNICEF immunization coverage working group. However, these can be improved upon to allow, for example, the exclusion of coverage data deemed implausible for estimation based on expert evaluation (unlikely zero estimates of coverage are excluded from the current analysis, although such estimates could sometimes reflect a vaccine stock-out). 
The methodology proposed in Section \ref{sec:method} can produce coverage estimates for all desired cases using available input data, although with greater uncertainty where input data are missing, hence offering the flexibility to use the most accurate input data for coverage estimation.    

\section{The proposed methodology}
\label{sec:method}
Our aim is to develop BHM approaches for producing estimates of national immunization coverage and associated uncertainties from multiple data sources. Here, we propose two candidate models termed balanced data single likelihood (BDSL) model and irregular data multiple likelihood (IDML) model. As the name implies, the BDSL model uses a single likelihood to capture the variability in the data and is considered here as a suitable alternative against which to compare the IDML model which induces considerable flexibility in the estimation of the variability in the data through using separate probability distributions.

In general,  let $\tilde{p}^{(k)}_{ijt}$ denote the $k$th type estimate of vaccination coverage (proportion) for the $i$th  country $i (i = 1, \dots, C)$, $j$th vaccine $j (j=1,...,V)$ and year $t (t=1, \ldots, T)$, where for 
\begin{itemize}
\item[(i)] $k=a$,   $\tilde{p}_{ijt}^{(a)}$ is  the administrative estimate,
\item[(ii)]  $k=o$,  $\tilde{p}_{ijt}^{(o)}$ is the official estimate,
\item[(iii)] $k=s$, $\tilde{p}_{ijt}^{(s)}$ is the  survey estimate.
 \end{itemize}
Figure~\ref{fig:multilikfig}
provides a plot of these estimates in percentage forms.

The three versions of the estimates $\tilde{p}_{ijt}^{(k)}$ are available to us, but the counts and the denominators corresponding to these estimated proportions are not always available for modelling. Hence we are not able to use the binomial distribution for modelling the true
vaccination coverage $p_{ijt}$.  Instead, we  treat the logit-transformed estimates,
$$
y^{(k)}_{ijt} \equiv \text{logit}\left(\tilde{p}_{ijt}^{(k)}\right)= \log\left( \frac{\tilde{p}_{ijt}^{(k)}}{1-\tilde{p}_{ijt}^{(k)}}\right),
$$
as observed data varying over the real line. Note that the  logit transformation, used here, is natural for transforming proportions to
the real line. As is well known, possible alternatives to the logit transformation are the inverse cumulative distribution function (cdf) transformation such as the probit, i.e.,$\tilde{p}_{ijt} = \Phi\left(y_{ijt} \right)$ where $\Phi(\cdot)$ is the cdf of the standard normal distribution. In this work, however, we only adopt the  logit transformation since it is able to accommodate more extreme values than the probit transformation.  
Hence $0 \le \tilde{p}_{ijt}  \le 1$ for all $i$, $j$ and $t$.

The transformed estimates are then assumed to follow the normal distribution based linear models. Indeed, see supplementary Figure 4 where the histograms of the logit and probit-transformed estimates (panels (b) and (c)) show better bell shaped curves compared to the histogram of the proportions (panel (a)) which is negatively skewed as expected. We also observe deviant peaks in the right tails of the histograms in panels (b) and (c), which is due to the high frequency of proportions close to 1 in the data.


Before we introduce the models we note that  although we model on the logit-transformed scale, we obtain and report the model based predictions on the original scale of 0 to 100\%. Sampling based Bayesian computation methods also allow us to obtain the uncertainties of the predictions on the original scale. Details to obtain these predictions  are provided in Section~\ref{sn:prediction}.

\subsection{Balanced data single likelihood (BDSL) model}
\label{sn:fulldatamodel}
  


We assume that all three estimates 
$y_{ijt}^{(a)}$,  $y_{ijt}^{(o)}$ and $y_{ijt}^{(s)}$ aim to estimate the true mean $\mu_{ijt}$ but each of these
three have their own biases. In the this model, these biases are captured using a source-specific random effect, $\nu^{(k)}$. The BDSL model attempts to model the observed $y^{(k)}_{ijt}$ as:
\begin{equation}\nonumber
  y^{(k)}_{ijt} = \mu_{ijt} +  \nu^{(k)} + e_{ijt}^{(k)}, 
\end{equation}
where
\begin{equation}
    \mu_{ijt} = \lambda + \beta_i + \alpha_j + \gamma_t + \phi_{it} + \delta_{jt} + \psi_{ij} + \omega_{ijt}.
    \label{eq:single1}
\end{equation} 
is the true bias-corrected mean, $e_{ijt}^{(k)}$ is an error term assumed to follow the
$N(0, \sigma^2)$ independently and identically for all values of $i, j$ and $t$, and $\lambda$ is the overall mean. Thus, the source-specific term, $\nu^{(k)}$, captures the bias of coverage estimates from data source $k$ relative to $\lambda$, and is modelled as $\boldnu=(\nu^{(a)}, \nu^{(o)}, \nu^{(s)})^{'} \sim N(0, \sigma^2_{\nu} \boldsymbol{I})$. Further, in equation (\ref{eq:single1}), $\beta_i$ is a country level effect, $\alpha_j$ is a vaccine effect and $\gamma_t$ is a temporal effect for $t = 1, \dots, T$, where $T$ denotes all time points being considered in the analysis. These terms capture overall variation in the data emanating from the different attributes. Additionally, $\phi_{it},\ \delta_{jt}, $ and $\psi_{ij}$ are country-time, vaccine-time and country-vaccine interaction terms modelling trends that are specific to each country $(\phi_{it})$ and vaccine $(\delta_{jt})$, and random variation between each country and vaccine $(\psi_{ij})$. Also, $\omega_{ijt}$ is a country-vaccine-time interaction that captures trends specific to each country-vaccine combination. Specification of these random effects are deferred to Section~\ref{sn:randomeff} below.   

We note that the source-specific random effect, $\nu^{(k)}$, is not included in the shared mean $\mu_{ijt}$ in (\ref{eq:single1}), as this only serves to estimate the biases of the data sources relative to $\mu_{ijt}$ and is hence undesirable. However, we note that model (\ref{eq:single1}) does not offer much flexibility to penalize the contribution of each data source to the shared mean since this is only achieved via the parameters $\nu^{(a)}, \nu^{(o)}$ and $\nu^{(s)}$, which are the same for all $i, j$ and $t$, and considering that $e_{ijt}^{(k)}$ is modelled as iid irrespective of $k$. Further discussions on this are provided in Section \ref{sec:idml}.


\subsection{Irregular data multiple likelihood (IDML) model}
\label{sec:idml}
The balanced model, as given in~(\ref{eq:single1}), is defined for all possible combinations of the indices  $i$, $j$, $k$ and $t$. Thus in reality, we need $3CVT$ data values where the factor 3 comes from three possible values of $k$, viz. admin, official and survey.  However, for MCV1 for example, we only have 65.1\% of these data values. Hence the remaining 34.9\% must be treated as missing in our Bayesian modelling.

\begin{figure}
    \centering
 \includegraphics[width=1.0 \textwidth]{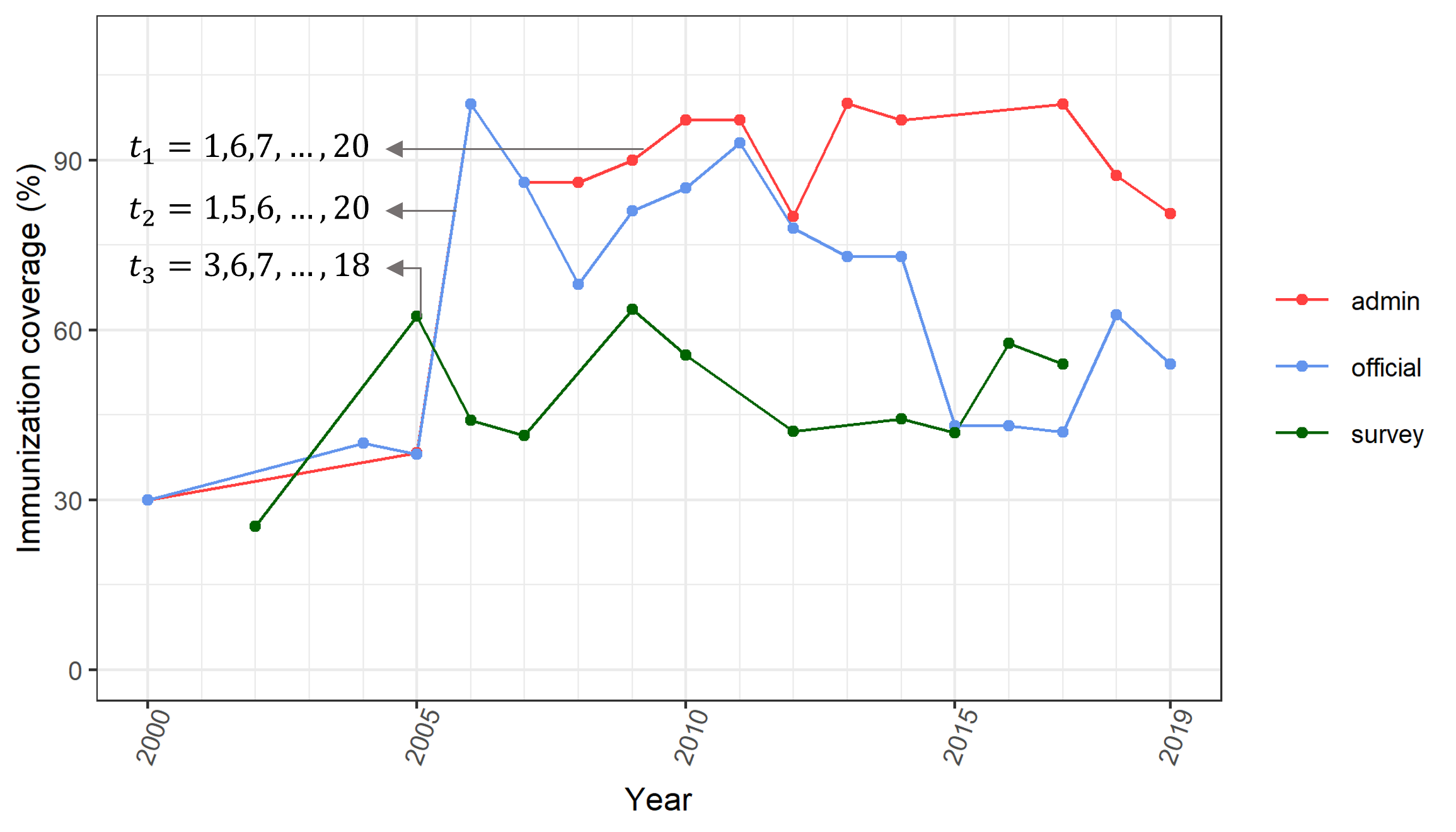} 
 \caption{Data illustration for the irregular data multiple likelihood model using estimates of MCV1 coverage for Nigeria.
\label{fig:multilikfig}} 
\end{figure}

The observed time points, denoted by the index $t$ in $\tilde{p}_{ijt}^{(k)}$, as seen in the
horizontal axis in Figure~\ref{fig:multilikfig}, are very much misaligned for the three types of estimates. In this figure,  the survey
and official estimates have been produced only for few selected years and not for all the years.
This presents a difficult problem in modelling based on balanced regular time series as for the BDSL model in Section~\ref{sn:fulldatamodel} since  all the data, $y_{ijt}^{(k)}$ are not available for all regularly spaced values of $t$ from 2000 to 2019. The problem is caused by the presence of a large percentage of missing data. Indeed, if we were to use regular time series models, then we will have 34.9\% of missing data for MCV1 from the three data series,  $y_{ijt}^{(k)}$, $k=a, o$ and $s$. A suitable multiple imputation scheme will be necessary to properly handle this large percentage of  missing data. 

Our proposed novel solution to this missing data problem comes through a multiple likelihood approach based on three different indices $t_1$, $t_2$, and $t_3$ respectively for admin, official and survey data respectively as illustrated in Figure~\ref{fig:multilikfig}. Thus, we model $y_{ijt_1}^{(a)}$,   $y_{ijt_2}^{(o)}$,  $y_{ijt_3}^{(s)}$ where the time 
indices $t_1$, $t_2$, and $t_3$ depend on the country $i$ and vaccine type $j$ combination. For example, in   Figure~\ref{fig:multilikfig} $t_1$ (for admin estimates) takes the values 1, 6, 7, $\ldots$, implying that the administrative estimates are missing for time points 2, 3, 4, and 5, and so on. In our modelling development, we simply write down the likelihood contributions based on the data from the observed time points. The combined likelihood function then captures all the information contained in the observed data for the underlying model parameters. 

The IDML model is given by:
\begin{eqnarray*}
  y_{ijt_1}^{(a)} 
  &= \lambda^{(a)} + \mu_{ijt_{1}} + \epsilon_{ijt_{1}},\ t_1 = 1, \dots, T_1, \\
  y_{ijt_2}^{(o)} 
  &= \lambda^{(o)} + \mu_{ijt_{2}} + \epsilon_{ijt_{2}},\ t_2 = 1, \dots, T_2,  \\
  y_{ijt_3}^{(s)} 
  &= \lambda^{(s)} + \mu_{ijt_{3}} + \epsilon_{ijt_{3}},\ t_3 = 1, \dots, T_3, 
\end{eqnarray*}
where 
\begin{equation} 
    \epsilon_{ijt_1} \sim N(0,  \sigma^2_1),\ \epsilon_{ijt_2} \sim N(0, \sigma^2_2),\ \epsilon_{ijt_3} \sim N(0, \sigma^2_3), 
\label{eq:eq1}
\end{equation}
and  $\sigma^2_k$ for $k=1, 2$ and 3 are source-specific error variance parameters.
Also, $\lambda^{(a)},\ \lambda^{(o)},\ $ and $\lambda^{(s)}$ are source-specific intercept terms capturing the bias of coverage estimates from data source $k$ relative to the true bias-corrected mean, $\mu_{ijt}$. The time indices  $t_1, t_2$ and $t_3$ are possibly unequally-spaced  denoting only the time points for which data are available from a given data source. Similarly, $T_1, T_2$ and $T_3$ are the total numbers of times data are available from the respective data sources. We note that if, for example, there are no survey data for country $i$ and vaccine $j$, then $T_3=0$.

The shared mean for model (\ref{eq:eq1}) is given by
\begin{equation}
    \mu_{ijt} = \beta_i + \alpha_j + \gamma_t + \phi_{it} + \delta_{jt} + \psi_{ij} + \omega_{ijt}, \quad t = 1,\dots,T.
\label{eq:eqb}
\end{equation}
Hence, the IDML model also brings out the  novel feature that to estimate $\mu_{ijt}$, we are directly able to combine information from all the available sources for that  specific $i$ (country), $j$ (vaccine) and $t$ (time), with appropriate relative weighting as estimated by the variance components $\sigma^2_1$,  $\sigma^2_2$ and  $\sigma^2_3$. Another likely advantage is that unlike the BDSL model, the IDML model does not need to estimate the input data for the missing cases, e.g. the admin estimate $y_{ijt}^{(a)}$ for MCV1 in Nigeria in 2006. Thus, the IDML model yields a lower dimensional parameter space which is advantageous in MCMC based Bayesian computation. 

We note that the shared mean in (\ref{eq:eqb}) does not include the source-specific intercept terms $\lambda^{(a)}$, $\lambda^{(o)}$ and $\lambda^{(s)}$. This is because these terms play a similar role as $\nu^{(k)}$ in the BDSL model in accounting for the biases arising from the various data sources and are also not desirable in the true, bias-corrected mean. However, unlike $\nu^{(k)}$ which is modelled as a random effect, these terms are modelled as fixed effects. Also, we note that unlike (\ref{eq:single1}) which includes an overall intercept term $\lambda$, the shared mean for the IDML model in (\ref{eq:eqb}) does not include an intercept term, which is a direct consequence of the different approaches adopted in accounting for the biases arising from the data sources in both models.  

In model (\ref{eq:eq1}), the data sources are weighted by their respective variance parameters -- $\sigma^2_1, \sigma^2_2,\ $ and $\sigma^2_3$, which in turn determine their contributions to the overall mean function in equation (\ref{eq:eqb}). The influence of each data source in the model can thus be controlled by adjusting the values of these parameters either directly or through the prior distributions placed on them. For example, to increase the influence of survey estimates in the model, an informative prior that encourages smaller values relative to $\sigma_1^2$ and $\sigma_2^2$ could be placed on $\sigma_3^2$. In contrast, the BDSL specification does not provide a mechanism for this adjustment since it assumes that $\sigma^2_1 = \sigma^2_2 = \sigma^2_3 = \sigma^2$. We investigated assigning separate prior distributions to $\nu^{(a)}$, $\nu^{(o)}$ and $\nu^{(s)}$, but this did not lead to any meaningful changes in the modelled estimates obtained from the shared mean in (\ref{eq:single1}). Hence, the IDML specification has more flexibility in terms of handling contributions from the data sources to the likelihood than the BDSL model.

It is straightforward to derive a country-level model from equations (\ref{eq:single1}) and (\ref{eq:eq1}) by dropping the $i$ subscript and excluding the terms: $\beta_i,\ \phi_{it},\ \psi_{ij}$ and $\omega_{ijt}$ from the model. Although such models can be implemented by individual countries, these are of less interest in the current work as they do not allow borrowing strength across countries.

\subsection{Specification of the random effects}
\label{sn:randomeff} 

We assume that the main effects $\beta_i$ and $\alpha_j$ are each iid normal random variables with mean zero and variances $\sigma_{\beta}^2$  and $\sigma_{\alpha}^2$, respectively. We note that it suffices to model country-level variation as random as previous work \citep{UtaziWHO2020} showed a lack of significant spatial dependence in the data. We assume a first-order autoregressive (AR(1)) model for the temporal effect $\gamma_t$. That is,
\begin{equation} 
  \gamma_t \sim N(\rho \gamma_{t-1}, \sigma^2_{\gamma}),
 \label{eq:gamdynamics}
\end{equation}
with $\gamma_1 \sim N(0, \sigma^2_{\gamma}/(1-\rho_{\gamma}^2))$. The country-time interaction effect
$$
\boldphi \sim N\left(\boldzero, [\sigma_{\phi}^{-2}\boldsymbol{Q}_{\phi}(\rho_{\phi})]^{-}\right),
$$
where 
$\boldphi = (\phi_{11}, \dots, \phi_{1T}, \dots, \phi_{C1}, \dots, \phi_{CT})$ and  $\boldsymbol{Q}_{\phi}(\rho_{\phi})$ is a $CT \times CT$ structured matrix \citep{Clayton1996,Knorr-Held2000} specifying the nature of interdependence between the elements of $\boldphi$, $\sigma_{\phi}^{-2}$ is an unknown precision parameter, $\rho_{\phi}$ is an autoregressive parameter and $[.]^-$ denotes the Moore-Penrose generalized inverse of a matrix. Following \cite{Clayton1996}, $\boldsymbol{Q}_{\phi}(\rho_{\phi})$ can be factorised as the Kronecker product of the structure matrices of the interacting main effects -- $\beta_i$ and $\gamma_t$. Given the nature of these terms, here we assume a Type II interaction \citep{Knorr-Held2000} such that $\boldsymbol{Q}_{\phi}(\rho_{\phi}) = \boldsymbol{I}_C \otimes \boldsymbol{R}_{\gamma}$, where $\boldsymbol{I}_C$ is an identity matrix and $\boldsymbol{R}_{\gamma}$ is the neighbourhood structure of an AR(1) process. This assumption implies that the temporal parameters for each country $\boldphi_{i,1:T} = (\phi_{i1}, \dots, \phi_{iT})$ follow an AR(1) process independent of all other countries. In other words, this two-way interaction term captures temporal trends that are different from country to country and do not have any spatial structure. Similarly, the vaccine-time interaction is modelled as 
$$\bolddelta  \sim N(\boldzero, [\sigma^{-2}_{\delta}\boldsymbol{Q}_{\delta}(\rho_{\delta})]^{-}),
$$
where $\bolddelta = (\delta_{11}, \dots, \delta_{1T}, \dots, \delta_{V1}, \dots, \delta_{VT})'$, $\sigma^{-2}_{\delta}$ is also an unknown precision parameter and $\rho_{\delta}$ is an autoregressive parameter. We also assume a Type II interaction for this parameter such that $\boldsymbol{Q}_{\delta}(\rho_{\delta}) = \boldsymbol{I}_V \otimes \boldsymbol{R}_{\gamma}$. Here again, the structure of $\boldsymbol{Q}_{\delta}$ implies that the temporal pattern for vaccine $j$ represented by $\bolddelta_{j,1:T} = (\delta_{j1}, \dots, \delta_{jT})$ is independent of those of other vaccines. Next, $\psi_{ij}$ represents the interaction of the iid terms $\beta_i$ and $\alpha_j$, hence we assume a Type I interaction \citep{Knorr-Held2000} for it, such that 
$$\boldpsi \sim N(\boldzero,\sigma^2_{\psi}\boldsymbol{I}_{CV}),
$$
where $\boldpsi = (\psi_{11}, \dots, \psi_{1V}, \dots, \psi_{C1}, \dots, \delta_{CV})'$ $\sigma_{\psi}^2$ is a variance parameter. This interaction term models additional unstructured country-vaccine variation. Lastly, the three-way interaction term, $\omega_{ijt}$, models temporal trends specific to each country-vaccine combination. We assume that
$$\boldomega \sim N(\boldzero, [\sigma^{-2}_{\omega}\boldsymbol{Q}_{\omega}(\rho_{\omega})]^{-}),
$$
where $\boldomega = (\omega_{111}, \dots, \omega_{CVT})'$ and all the parameters are as defined previously. We also assume a Type-II-like interaction for $\boldomega$ such that $\boldsymbol{Q}_{\omega}(\rho_{\omega}) = \boldsymbol{I}_C \otimes \boldsymbol{I}_V \otimes \boldsymbol{R}_{\gamma} = \boldsymbol{I}_{CV} \otimes \boldsymbol{R}_{\gamma}$, which implies that the temporal structure given to country $i$ and vaccine $j$, represented by $\boldomega_{ij, 1:T} = (\omega_{ij1}, \dots, \omega{ijT})$, is independent of those of other country-vaccine combinations.


\section{Bayesian inference and computation}

In this section, we describe details of implementation of the proposed models in a Bayesian setting. 
Let $\boldy$  denote all observed data. Also, let
\begin{eqnarray*}
\boldy^{(a)} & = &  \big( y_{111}^{(a)},  \dots, y_{CVT_{1}}^{(a)} \big)', \\
\boldy^{(o)} & = & \big( y_{111}^{(o)},  \dots, y_{CVT_{2}}^{(o)} \big)' \quad  \text{ and} \\ 
\boldy^{(s)} & = & \big( y_{111}^{(s)},  \dots, y_{CVT_{3}}^{(s)} \big)',
\end{eqnarray*}
where $T_1=T_2=T_3=T$ in model (\ref{eq:single1}). For the BDSL model, let $\boldeta^{B} = (\lambda,\ \boldbeta,\ \boldalpha,\ \boldgamma$, $\ \boldphi,\ \bolddelta,\ \boldpsi,\ \boldomega, \boldnu)$ denote a latent field comprising the intercept term and the joint distribution of all the parameters in the mean model $\boldmu = (\mu_{111}, \dots, \mu_{CVT})$ (i.e., the random effects) given in equation (\ref{eq:single1}), $\theta_1^{B} = \sigma^2$ denote the variance of the Gaussian observations, and
$$
\boldtheta_2^{B} = (\sigma_{\beta}^2,\ \sigma_{\alpha}^2,\ \rho_{\gamma},\ \sigma^2_{\rho}, \ \rho_{\phi}, \ \sigma^2_{\phi},\ \rho_{\delta},\ \sigma^2_{\delta}, \ \sigma^2_{\psi},\ \rho_{\omega},\ \sigma^2_{\omega}, \sigma^2_\nu)'
$$
denote the  hyperparameters of the latent field $\boldeta^{B}$, i.e. the variances and autocorrelation parameters of the random effects. Similarly, for the IDML model, let 
\begin{eqnarray*}
\boldeta^{I} &=& (\lambda^{(a)},\ \lambda^{(o)},\ \lambda^{(s)},\ \boldbeta,\ \boldalpha,\ \boldgamma,\ \boldphi,\ \bolddelta,\ \boldpsi,\ \boldomega),\\
\boldtheta_1^{I} &=& (\sigma_1^2,\ \sigma_2^2,\ \sigma_3^2)' \quad \text{and} \\
\boldtheta_2^{I} &=& (\sigma_{\beta}^2,\ \sigma_{\alpha}^2,\ \rho_{\gamma},\ \sigma^2_{\rho},\ \rho_{\phi},\ \sigma^2_{\phi},\ \rho_{\delta},\ \sigma^2_{\delta},\ \sigma^2_{\psi},\ \rho_{\omega},\ \sigma^2_{\omega})'.
\end{eqnarray*}

We complete our Bayesian specification by placing appropriate prior distributions on the parameters as follows. For the BDSL model,
we assume the following prior distributions: 
\begin{eqnarray*}
  \lambda & \sim &  N(0,1), \\
  \sigma & \sim &  \text{Cauchy}(0,2) I(\sigma > 0), \\
  \sigma_\nu & \sim &  \text{Cauchy}(0,2) I(\sigma_\nu > 0).
\end{eqnarray*}
For the IDML model, the prior distributions were:
$$
  \lambda^{(a)}  \sim   N(0, v_1); \quad 
  \lambda^{(o)}  \sim   N(0, v_2); \quad 
  \lambda^{(s)}  \sim   N(0, v_3);
  $$
 \vspace*{-10mm}
 \begin{eqnarray*}
  \sigma_1  & \sim &   \text{Cauchy}(0,2) I(\sigma_1 >0); \\
  \sigma_2  & \sim &   \text{Cauchy}(0,2) I(\sigma_2 >0);  \\ 
  \sigma_3  & \sim &   \text{Cauchy}(0,0.2) I(0\leq \sigma_3 \leq 0.4).
\end{eqnarray*}
These prior distributions were chosen based on trial runs, during which we set $v_1 = v_2 = v_3 = 0.25$. The highly informative truncated Half-Cauchy prior on $\sigma_3$ was chosen to attribute greater likelihood to survey estimates in the model subject to expert belief and to adjust for the higher proportions of missingness in this data source. For all other parameters in both models, we used default uniform priors available in Stan \cite{stan}. 

Letting $\boldtheta^{B} = (\theta_1^{B}, \boldtheta_2^{B})'$, the joint posterior distribution of the BDSL model can be written as:
\begin{align}
    \pi(\boldtheta^{B}, \boldeta^{B} | \boldy) &\propto p(\boldy|\boldeta^{B},\theta_1^{B}) \times p(\boldeta|\boldtheta_2^{B}) \times p(\boldtheta^{B}), \nonumber \\
    &\propto p(\boldy^{(a)}|\boldeta^{B}, \sigma^2) \times p(\boldy^{(o)}|\boldeta^{B}, \sigma^2) \times p(\boldy^{(s)}|\boldeta^{B}, \sigma^2) \times p(\boldbeta|\boldtheta_2^{B}) \times p(\boldalpha|\boldtheta_2^{B}) \nonumber \\ 
    &\times p(\boldgamma|\boldtheta_2^{B}) \times p(\boldphi|\boldtheta_2^{B}) \times p(\bolddelta|\boldtheta_2^{B}) \times p(\boldpsi|\boldtheta_2^{B}) \times p(\boldomega|\boldtheta_2^{B}) \times p(\boldnu|\boldtheta_2^{B}) \times p(\boldtheta^{B}),\nonumber\\
    &\propto \prod^C_{i=1} \prod^V_{j=1} \prod^T_{t=1} \prod^3_{k=1} \bigg[ \sigma^{-1} \exp\bigg( -\frac{1}{2\sigma^2_1}(y_{ijt}^{(k)} - \mu_{ijt} - \nu^{(k)})^2 \bigg)\bigg] \nonumber \\
    &\times \prod_{i=1}^{C} \sigma_{\beta}^{-1} \exp\Bigg( -\frac{\beta_i^2}{2\sigma^2_{\beta}} \Bigg) \times \prod_{j=1}^{V} \sigma_{\alpha}^{-1} \exp\Bigg( -\frac{\alpha_j^2}{2\alpha_{\alpha}^2} \Bigg) \nonumber \\
    &\times \sqrt{\sigma_{\gamma}^{-2}( 1 - \rho_{\gamma}^2 )} \exp \bigg( -\frac{(1-\rho_{\gamma}^2)}{2\sigma^2_{\gamma}}\gamma_1^2 \bigg) \times \prod_{t=2}^{T}\sigma_{\gamma}^{-1} \exp\bigg( -\frac{(\gamma_t - \rho\gamma_{t-1})^2}{2\sigma_{\gamma}^2} \bigg) \nonumber \\
    &\times |\sigma^{-2}_{\phi} \boldsymbol{Q}_{\phi}(\rho_{\phi})|^{\frac{1}{2}} \exp \bigg( -\frac{1}{2\sigma^2_{\phi}}\boldphi^{'} \boldsymbol{Q}_{\phi}(\rho_{\phi}) \boldphi \bigg) \nonumber \\
    &\times |\sigma_{\delta}^{-2} \boldsymbol{Q}_{\delta}(\rho_{\delta})|^{\frac{1}{2}} \exp \bigg( -\frac{1}{2\sigma^2_{\delta}}\bolddelta^{'} \boldsymbol{Q}_{\delta}(\rho_{\delta})\bolddelta \bigg) \times \sigma_{\psi}^{-1} \exp\bigg( -\frac{1}{2\sigma_{\psi}^2} \boldpsi^{'} \boldpsi \bigg) \nonumber \\
    &\times |\sigma_{\omega}^{-2}\boldsymbol{Q}_{\omega}(\rho_{\omega})|^{\frac{1}{2}} \exp \bigg( -\frac{1}{2\sigma^2_{\omega}}\boldomega^{'} \boldsymbol{Q}_{\omega}(\rho_{\omega})\boldomega \bigg) \times \sigma_{\nu}^{-1} \exp\bigg( -\frac{1}{2\sigma^2_{\nu}} \boldsymbol{\nu}^{'} \boldsymbol{\nu} \bigg) \nonumber\\
    &\times p(\boldtheta^B),
\label{eq:eq6}
\end{align}
where $p(\boldtheta)$ denotes the joint prior distribution on the parameters. Given that we assumed that the parameters are $a priori$ independent,  $p(\boldtheta^B)$ simply represents the product of the prior distributions assigned to them. 

Similarly, letting $\boldtheta^I = (\boldtheta_1^I, \boldtheta_2^I)'$, the joint posterior distribution of the IDML model can be written as: 
\begin{align}
    \pi(\boldtheta^I,\boldeta^I|\boldy) &\propto p(\boldy|\boldeta^I, \boldtheta_1^I) \times p(\boldeta^I|\boldtheta_2^I) \times p(\boldtheta^I), \nonumber \\
     &\propto p(\boldy^{(a)}|\boldeta^I, \boldtheta_1^I) \times p(\boldy^{(o)}|\boldeta^I, \boldtheta_1^I) \times p(\boldy^{(s)}|\boldeta^I, \boldtheta_1^I) \times p(\boldsymbol{\beta}, \boldsymbol{\alpha}, \boldsymbol{\gamma}, \boldsymbol{\phi}, \boldsymbol{\delta}, \boldsymbol{\psi}, \boldsymbol{\omega} | \boldtheta_2^I) \nonumber \\
    &\times p(\boldtheta^I), \nonumber\\
    &\propto \prod_{i=1}^C \prod_{j=1}^V \left[ \prod_{t_{1} = 1}^{T_{1}} \sigma_1^{-1} \exp\Bigg( -\frac{1}{2\sigma_1^2}\big( y_{ijt_{1}}^{(a)} - \lambda^{(a)} - \mu_{ijt_{1}} \big)^2 \Bigg)\right. \nonumber \\
    &\times \prod_{t_{2} = 1}^{T_{2}} \sigma_2^{-1} \exp\Bigg( -\frac{1}{2\sigma_2^2}\big( y_{ijt_{2}}^{(o)} - \lambda^{(o)} - \mu_{ijt_{2}} \big)^2 \Bigg) \nonumber \\
    &\left. \times \prod_{t_{3} = 1}^{T_{3}} \sigma_3^{-1} \exp\Bigg( -\frac{1}{2\sigma_3^2}\big( y_{ijt_{3}}^{(s)} - \lambda^{(s)} - \mu_{ijt_{3}} \big)^2 \Bigg)\right] \nonumber \\
    & \times p(\boldsymbol{\beta}, \boldsymbol{\alpha}, \boldsymbol{\gamma}, \boldsymbol{\phi}, \boldsymbol{\delta}, \boldsymbol{\psi}, \boldsymbol{\omega} | \boldtheta_2^I) \times p(\boldtheta^I),
\label{eq:eq5}
\end{align}
where $p(\boldbeta, \boldalpha, \boldgamma, \boldphi, \bolddelta, \boldpsi, \boldomega | \boldtheta_2^I) = p(\boldbeta|\boldtheta_2^I) \times p(\boldalpha|\boldtheta_2^I) \times p(\boldgamma|\boldtheta_2^I) \times p(\boldphi|\boldtheta_2^I) \times p(\bolddelta|\boldtheta_2^I) \times p(\boldpsi|\boldtheta_2^I) \times p(\boldomega|\boldtheta_2^I)$, which are all the same as corresponding expressions provided in equation (\ref{eq:eq6}). 

The goal of inference in equations (\ref{eq:eq6}) and (\ref{eq:eq5}) is to estimate the posterior distributions of the components of $\boldeta^B$, $\boldeta^I$, $\boldtheta^B$ and $\boldtheta^I$, both of which are in turn used to obtain the underlying, smoothed coverage estimates $\mu_{ijt},\ \forall\ i \in \{ 1, \dots, C \},\ j \in \{ 1, \dots, V \} \text{and}\ t \in \{ 1, \dots, T \}$ as given in equations (\ref{eq:single1}) and (\ref{eq:eqb}).

We fitted both models by running Markov Chain Monte Carlo (MCMC) using the NUTS (No-U-Turn Sampler) algorithm \citep{Hoffman_Gelman_2014} within the Stan package in R \cite{rstan}. We implemented four chains, each of which was run for 4,000 iterations including a burn-in of 2,000 iterations. We assessed convergence using the MCMC convergence statistic, $\hat{R}$, which we ensured was below 1.05 \citep{Vehtarietal2021} for each parameter in the models. We also provide an R package for implementing the proposed models, further details of which are provided in Section \ref{sec:software}.

\subsection{Smoothed overall estimates}
\label{sn:smoothoverall}

Note that the mean models in (\ref{eq:single1}) and ~(\ref{eq:eqb}) are well defined for both the BDSL and IDML models as discussed above.
We use the posterior distribution of   $\mu_{ijt}$ to produce our model based vaccination coverage estimates.
After eliminating the source-specific bias components, $\nu^{(k)}$ and $\lambda^{(k)}$ in our notation, we estimate the coverage estimate as follows.
We suppose that the inverse logit-transform 
\begin{equation}
  p_{ijt} = \frac{1}{1+ \exp\left(-\mu_{ijt}\right)}
  \label{eq:invlogit}
\end{equation}
is the source free and true bias-corrected vaccination coverage proportion. 

The posterior distribution of $p_{ijt}$ given all the data
is summarised to provide source free estimates of  vaccination coverage. The posterior distribution of $p_{ijt}$ is easy to
calculate using MCMC sampling. For example, we obtain MCMC samples $\boldtheta^{(\ell)}$, $\ell = 1, \ldots, L$ for all the unknown parameters and
random effects and missing data collectively  denoted by $\boldtheta$. Using $\boldtheta^{(\ell)}$ we evaluate $\mu_{ijt}^{(\ell)}$ and subsequently 
$p_{ijt}^{(\ell)}$ for $\ell = 1, \ldots, L.$ These samples are then used to estimate the true  $p_{ijt}$ along with the uncertainty estimates.
Note that these uncertainty estimates are obtained exactly for vaccination coverage at the original scale.

\subsection{Prediction and aggregation to the regional level}
\label{sn:prediction}

In addition to parameter estimation, it is often the goal in Bayesian analysis to estimate missing observations or to predict future observations. Typically, Bayesian in-sample and out-of-sample prediction (the latter is also known as forecasting) are both based on the posterior predictive distribution. For example, the one-step-ahead 
prediction at any time point $t$ can be obtained by evaluating the conditional distribution of $p_{ijt+1}$ given all the data $\boldy$.
According to~(\ref{eq:invlogit}) we have:
$$
p_{ijt+1} = \frac{1}{1+ \exp\left(-\mu_{ijt+1}\right)}, 
$$
where, for example, for the BDSL model, 
$$
\mu_{ijt+1} = \lambda + \beta_i + \alpha_j + \gamma_{t+1} + \phi_{it+1} + \delta_{jt+1} + \psi_{ij} + \omega_{ijt+1}
$$
from~(\ref{eq:single1}). Hence to predict $p_{ijt+1}$ we also need the values of the time advanced parameters  $\gamma_{t+1}$,  $\delta_{jt+1}$ and
$\omega_{ijt+1}$ at time $t$. For a given $t$ within the modelled time period $T$, these parameters are already sampled within the implemented
MCMC scheme. For $t \ge T$ we use the assumed model dynamics, e.g. (\ref{eq:gamdynamics}) to sample these parameters.
That is, we set
$$
\gamma_{t+1}^{(\ell)} \sim N(\rho^{(\ell)} \gamma_{t}^{(\ell)}, \sigma^{2(\ell)}_{\gamma}),
$$
if $\gamma_{t+1}$ has not been  sampled already. The other dynamic parameters are treated similarly. 
Hence to estimate (or to predict if $t>T$) $p_{ijt}$,  using either of the proposed models, we evaluate the posterior distribution of
$p_{ijt}$ given $\boldy$ by drawing samples $p_{ijt}^{(\ell)}$ for $\ell=1, \ldots, L$ for a large number of MCMC replicates $L$. 

In Stan, these out-of-sample predictions can be computed post model-fitting. As with in-sample estimation, the final predictions can be obtained by summarizing the inverse logit of the posterior samples of $\mu_{ij(t+1)}^{(\ell)}$. We note that by default, in-sample estimates of $\mu_{ijt}$ are estimated for desired in-sample country-vaccine-time combinations even when no data are observed for these cases, since $\mu_{ijt}$ is estimated $\forall i \in \{1, \dots, C\},\ j \in \{1, \dots, V\},\ t \in \{1, \dots, T \}$. As explained earlier, in-sample estimates of $\mu_{ijt}$ are processed post-model-fitting using year of vaccine introduction (yovi) data to obtained modelled estimates for desired country-vaccine-time combinations.

Further, we obtain modelled estimates  of immunization coverage for each WHO region as population-weighted averages taken over all the countries falling within the region. That is, for region $R_r\ (r = 1, \dots, 6)$, vaccine $j$, time $t$ and posterior sample $\ell$,
\begin{equation}
    p_{rjt}^{(\ell)} (R) = \sum_{i\in R_r} p_{ijt}^{(\ell)} \times q_i^r,
\label{eq:eq8}
\end{equation}
where $q_i^r$ is the proportion of surviving infants or target population for MCV2 of region $R_r$ living in country $i$. It is straightforward to compute equation (\ref{eq:eq8}) using the posterior distributions of $p_{ijt}$ under each model. 

\subsection{Model comparison, evaluation and validation}

To choose between the proposed models in our application, we considered the Watanabe-Akaike information criterion (WAIC) \citep{watanabe2013}. The WAIC is a fully Bayesian criterion that is based on the log of the predictive density for each data point, hence it assesses the ability of the fitted model to predict the input data. Accessible discussions regarding  WAIC are provided by \cite{gelman_waic2014} and \cite{Sahubook}. 

To further evaluate the ability of the proposed models to predict the in-sample and out-of-sample data (i.e. $p_{ijt}$) in a simulation experiment (due to lack of true values of $p_{ijt}$ in our application - see Section \ref{sec:simstudy}), we computed the following metrics:
\begin{align*}
    \textrm{Average bias: } AvBias &= \frac{1}{m}\sum^m_{k=1}(\hat{p}_k - p_k); \\
    \textrm{Root mean squared error: } RMSE &= \sqrt{\frac{1}{m}\sum^m_{k=1} (\hat{p}_k - p_k)^2}; \\
    \textrm{Mean absolute error: } MAE &= \frac{1}{m}\sum^m_{k=1}|\hat{p}_k - p_k|; \\
    \textrm{95\% coverage: } \text{95\% coverage} &= 100 \times \sum^m_{k=1} I(\hat{p}^l_k \leq p_k \leq \hat{p}_k^u);
\end{align*}
and the Pearson's correlation between observed and predicted values. Here, $m$ denotes all the values of $p_{(.)}$ used for validation (across all vaccines, countries and time points), $\hat{p}_k$ and $p_k$ are the predicted (i.e. the posterior means) and observed values,  $\hat{p}_k^l$ and $\hat{p}_k^u$ are the lower and upper limits of the 95\% credible intervals of the predictions and $I(.)$ is an indicator function.  The actual coverage of the 95\% prediction intervals assesses the accuracy of the uncertainty estimates associated with the predictions, while all the other metrics evaluate the accuracy of the point estimates. The closer the 95\% coverage rates are to the nominal value of 95\%, the better the predictions. Similarly, the closer the RMSE, MAE and AvBias (in absolute value) are to zero, the better the predictions. Correlations close to 1 indicate strong predictive power. 

\section{Simulation study} 
\label{sec:simstudy}

We conducted a simulation study to examine the predictive performance of the proposed models with respect to in-sample and out-of-sample estimation of $p_{ijt}$. In the study, we set $C=20, T=20$ and $V=5$, mimicking a moderately-sized WHO region. We then used the following true parameter values to generate estimates of $\mu_{ijt}$, as described in equations (\ref{eq:single1}) and (\ref{eq:eqb}) for the BDSL and IDML models, respectively: $\sigma^2_{\beta} = 1,\ \sigma^2_{\alpha} = 1,\ \rho_{\gamma} = 0.5,\ \sigma^2_{\gamma} = 1,\ \rho_{\phi} = 0.3,\ \sigma^2_{\phi} = 0.25,\ \rho_{\delta} = 0.4, \sigma^2_\delta = 0.64,\ \sigma^2_{\psi} = 1,\ \rho_{\omega} = 0.7,\ \sigma^2_{\omega} = 0.64$. Additionally, for the IDML model, we set $\lambda^{(a)} = 0.07, \ \lambda^{(o)} = 0.02,\ \lambda^{(s)} = 0.05$. For the BDSL model, we set $\sigma^2=1$ and $\lambda$ equal to the means of the corresponding parameters in the IDML model, i.e. $\lambda = 0.05$.

Owing to the key role that the parameters $\sigma_1^2,\ \sigma_2^2,\ \sigma_3^2$ and $\sigma_\nu^2$ play in capturing the amount of residual variability or bias attributable to the different data sources in the proposed models, we examined the effect of their varying values on the estimation of $p_{ijt}$ (equation (\ref{eq:invlogit})) by considering the following scenarios.\\
\noindent\\
\noindent \textbf{Scenario 1:} Variance of $\nu^{(k)} (\text{i.e. } \sigma_\nu^2)$ in BDSL model set equal to the average of the conditional variances of admin ($\sigma_1^2$), official ($\sigma_2^2$) and survey estimates ($\sigma_3^2$) in IDML model ($\sigma_1^2 = 1,\ \sigma_2^2 = 0.64,\ \sigma_3^2 = 0.16$ and $\sigma_\nu^2 = 0.6$).\\
\noindent\\
\noindent \textbf{Scenario 2:} Large differences between the conditional variances of admin, official and survey estimates and larger variance for $\nu^{(k)} (\sigma_1^2 = 9,\ \sigma_2^2 = 4,\ \sigma_3^2 = 0.25$ and $\sigma_\nu^2 = 4$).\\
\noindent\\
\noindent \textbf{Scenario 3:} Same conditional variances for admin, official and survey estimates and smaller variance for $\nu^{(k)}\ (\sigma_1^2 = 1,\ \sigma_2^2 = 1,\  \sigma_3^2 = 1$ and $\sigma_\nu^2 = 0.1)$.\\
\noindent\\
These true parameter estimates were chosen to encourage, as much as possible, an even distribution of values of $p_{ijt}$ on the unit interval. Adding the other components of equations (\ref{eq:single1}) and (\ref{eq:eq1}) to the simulated values of $\mu_{ijt}$, we obtained the simulated admin, official and survey estimates for all values of $C$, $T$ and $V$. To mimic the patterns of missingness in MCV2 and PCV3 in our application (see, e. g., supplementary Figure 3), we randomly selected $t = 10$ or $t = 15$ as the starting points for the observations for the last two vaccines. Further, we deleted 15\% of each of the simulated admin and official data and 20\% of the simulated survey data to reflect the overall pattern of missing values in our application. In all, for each model, the simulated data had a total of 3864 observations, 68\% of which were either admin or official data, while the remaining 32\% were survey data. 

We placed the same prior distributions as before on both models, except that for the IDML model, we placed a Half-Cauchy$(0,2)$ prior on $\sigma_3$, making it the same as the priors on $\sigma_1$ and $\sigma_2$, since the goal here is to compare both models under similar conditions (the effect of the prior specification on $\sigma_3$ on the performance of the IDML model is examined further in Section \ref{sec:application}). We set the starting point for the one- and two-step ahead predictions at $t=11$, meaning that we would use the first 10 observations as base years and then make predictions for the remaining ten time points, i.e. $t=11,\dots,20$.

\begin{table}[h]
\centering
\begin{tabular}{ l  r  r  r  r  r  r }
    \hline \hline \hline
    {\bf Validation} & \multicolumn{6}{c}{\bf In-sample prediction} \\
    \cline{2-7}
    {\bf metrics} & \multicolumn{2}{c}{\bf Scenario 1} & \multicolumn{2}{c}{\bf Scenario 2} &  \multicolumn{2}{c}{\bf Scenario 3} \\
    \cline{2-7}
    & {\bf BDSL} & {\bf IDML} &  {\bf BDSL} & {\bf IDML} &  {\bf BDSL} & {\bf IDML} \\
    \hline
    {AvBias} & -8.81 & {\bf 0.79} & -9.25  & {\bf 0.51} & {\bf -1.00} & 1.12  \\
    {RMSE} & 11.22 & {\bf 3.47}  & 11.95 & {\bf 3.87} & 5.74 & {\bf 4.98}  \\
    {MAE} & 9.09 & {\bf 1.73}  & 9.50 & {\bf 1.87}  & 3.78 & {\bf 2.92}  \\
    {95\% coverage} & 79.00 & {\bf 98.50}  & 28.20 & {\bf 98.40}  & 74.10 & {\bf 98.00}  \\
    {Correlation }& 0.98 & {\bf 0.99} & 0.97 & {\bf 0.99}  & 0.98 & {\bf 0.99} \\
    \hline 
    & \multicolumn{6}{c}{\bf One-step-ahead prediction} \\
    {AvBias} & -3.53 & {\bf -2.11}  & {\bf -1.58} & -2.54  & {\bf -0.21} & -1.82  \\
    {RMSE} & 21.40 & {\bf 18.60}  & 19.40 & {\bf 18.55} & 20.74 & {\bf 19.10} \\
    {MAE} & 18.54 & {\bf 15.82} & 15.89 & {\bf 15.74} & 17.29  & {\bf 16.20} \\
    {95\% coverage} & {\bf 99.00} & 99.70 & {\bf 98.70} & 99.80 & {\bf 98.20} & 99.70  \\
    {Correlation} & 0.74 & {\bf 0.81} & 0.79 & {\bf 0.81} & 0.75  & {\bf 0.80} \\
    \hline 
    & \multicolumn{6}{c}{\bf Two-step-ahead prediction} \\
    {AvBias} & -3.33 & {\bf -2.78} & {\bf -2.20} & -3.19 & {\bf -1.23} & -2.49  \\
    {RMSE} & 21.38 & {\bf 19.42} & 20.10 & {\bf 19.37} & 21.34 & {\bf 19.87}  \\
    {MAE} & 18.36  & {\bf 16.70} & 16.68 & {\bf 16.64} & 18.01 & {\bf 17.00}  \\
    {95\% coverage} & {\bf 98.79} & 99.74 & {\bf 98.58} & 99.79 & {\bf 98.47} & 99.74  \\
    {Correlation} & 0.74 & {\bf 0.79} & 0.77  & {\bf 0.79} & 0.73 & {\bf 0.78} \\
    \hline \hline \hline
\end{tabular}
\caption{Simulation study: Validation statistics for in-sample, one-step-ahead and two-step ahead predictions. The better result is shown in bold in each case.\vspace{0.2cm}}
\label{tab:simresults}
\end{table}

In Table \ref{tab:simresults}, we report the results of the study showing the validation statistics computed using the true and modelled estimates of $p_{ijt}$ under each model. Also, in supplementary Figure 5, we show examples of the simulated data and modelled estimates for five countries and three vaccines under scenario 1. For in-sample prediction, the IDML model generally outperformed the BDSL model across the three scenarios, yielding both more accurate point and uncertainty estimates. In particular, we observe that the BDSL model had relatively large AvBias estimates under Scenarios 1 and 2, demonstrating that it under-predicted the true values of $p_{ijt}$ in those cases (see, e.g., supplementray Figure 5). This under-prediction also resulted in very poor 95\% coverage values under Scenario 2. This suggests that the BDSL model may not be well-suited for in-sample prediction when there is considerable amount of variation (or bias) arising from the different data sources.

For out-of-sample prediction, the RMSE, MAE and correlation estimates are worse off as expected. Mixed results were, however, obtained when considering AvBias and 95\% coverage. We note that the high values of the actual 95\% coverage for both models are not surprising since out-of-sample predictions are often made with wider uncertainty intervals and the goal here was to predict the true values of $p_{ijt}$ and not the random observations. We rather focus on using the other metrics to evaluate the out-of-sample performance of the models. Again, the IDML model generally yielded more accurate estimates of $p_{ijt}$ than the BDSL model according to the RMSE, MAE and correlation, although it tended to produce relatively more biased estimates, especially under scenarios 1 and 2. Unlike in in-sample prediction, the performance of the BDSL model appears to be more stable and more comparable to that of the IDML model across the three scenarios in out-of-sample prediction. In all, these results show that the IDML model outperformed the BDSL model and is, therefore, better suited for both in- and out-of-sample prediction with regards to the estimation of the true, underlying coverage estimates, $p_{ijt}$.

\section{Results}
\label{sec:application}

Here, we present and discuss the results of the application of the proposed methodology to produce modelled estimates of national immunization coverage for all WHO Member States.  

\subsection{Model choice, validation and parameter estimates}
We first fitted both the BDSL and IDML models to the national immunization coverage data to further examine their performance and suitability for the data. With the IDML model, we considered two cases - one in which we placed a Half-Cauchy$(0, 2)$ prior on $\sigma_3$  to depict an unrestricted scenario, and the other where we placed a truncated Half-Cauchy$(0, 0.2)$ prior on $\sigma_3$ to improve the contribution of survey data to the likelihood. For the BDSL model, we used the same priors described previously. 

\begin{table}[ht]
\centering
\begin{tabular}{ l  r  r  r  r  r  r }
    \hline \hline \hline
    {\bf Region} & \multicolumn{3}{c}{\bf BDSL} & \multicolumn{3}{c}{\bf IDML} \\ [0.5 ex]
    \cline{2-7}
    & {\bf GOF} & {\bf penalty} & {\bf WAIC} & {\bf GOF} & {\bf penalty} & {\bf WAIC} \\ 
    \hline
    \multicolumn{7}{c}{Half-Cauchy$(0,2)$ prior placed on $\sigma_3$ in the IDML model} \\ [0.5 ex]
    \hline 
    AFR  & 21524.8 & {\bf 1053.9} & {\bf 23632.6} & {\bf 20905.1} & 1543.5 & 23992.1 \\
    AMR  & 8024.3 & {\bf 1314.8} & 10653.9 & {\bf 1370.3} & 2242.1 & {\bf 5854.5} \\
    EMR  & 6905.6 & {\bf 733.0} & 8371.6 & {\bf 5699.9} & 1093.4 & {\bf 7886.7}\\
    EUR & 7990.8 & {\bf 1453.4} & 10897.6 & {\bf 5416.9} & 2325.0 & {\bf 10066.9} \\
    SEAR  & 5094.2 & {\bf 189.4} & 5473.0 & {\bf 4892.6} & 269.6 & {\bf 5431.8}\\
    WPR  & 8656.1 & {\bf 850.7} & 10357.5 & {\bf 7276.0} & 1302.8 & {\bf 9881.6} \\
    \hline
    \multicolumn{7}{c}{Truncated Half-Cauchy$(0,0.2)$ prior placed on $\sigma_3$ in the IDML model} \\ [0.5 ex]
    \hline 
    AFR & {\bf 21524.8} & {\bf 1053.9} & {\bf 23632.6} & 22530.8 & 1420.1 & 25371.0 \\
    AMR & {\bf 6163.8} & 2599.3 & 11362.4 & 8024.3 & {\bf 1314.8} & {\bf 10653.9} \\
    EMR & {\bf 6905.6} & {\bf 733.0} & {\bf 8371.6} & 7019.2 & 1113.2 & 9245.6 \\
    EUR & 7990.8 & {\bf 1453.4} & {\bf 10897.6} & {\bf 6779.1} & 2438.1 & 11655.3 \\
    SEAR & 5094.2 & {\bf 189.4} & {\bf 5473.0} & {\bf 4963.7} & 357.8 & 5679.3 \\
    WPR & {\bf 8656.1} & {\bf 850.7} & {\bf 10357.5} & 9141.6 & 1237.8 & 11617.2 \\
    \hline \hline \hline
\end{tabular}
\caption{WAIC statistics for the balanced data single likelihood (BDSL) and irregular data multiple likelihood (IDML) models for each WHO region. The better result is shown in bold in each case.\vspace{0.2cm}}
\label{tab:tap1}
\end{table}

In Table \ref{tab:tap1}, we report the WAIC statistics for these models, which reveal an interesting pattern. With a less restrictive Half-Cauchy$(0, 2)$ prior on $\sigma_3$, the IDML model clearly outperformed the BDSL model in all cases except in the AFR region. When examining the contributions of the penalty and goodness-of-fit (GOF) terms to the calculated WAIC, we observe that although the BDSL model has smaller penalties as expected, since it includes a fewer number of parameters, the IDML model consistently provided better fits according to the GOF statistics. However, the use of a truncated Half-Cauchy$(0, 0.2)$ prior on $\sigma_3$  in the IDML model, though deliberate, resulted in poorer fits to the data since this prior biases the modelled estimates towards survey data. Hence, the BDSL model yielded smaller WAIC statistics in this case. These results further demonstrate the flexibility of the IDML model to adjust the modelled estimates based on expert opinions. All modelled outputs and results presented in the remainder of this work are, therefore, based on the IDML model only. 

\begin{table}[ht]
\centering
\begin{tabular}{ c  r  r  r  r  r }  
    \hline \hline \hline
    \textbf{Parameter} & \textbf{Mean} & \textbf{Std. dev.} & \textbf{2.5\%} & \textbf{50\%} & \textbf{97.5\%} \\ [0.5ex]
    \hline 
    $\hat{\lambda}^{(a)}$ & 0.4922 & 0.2868 & -0.0728 & 0.4887 & 1.0557 \\
    $\hat{\lambda}^{(o)}$ & 0.3661 & 0.2870 & -0.2023 & 0.3656 & 0.9286 \\
    $\hat{\lambda}^{(s)}$ & -0.421 & 0.2869 & -1.0068 & -0.4421 & 0.1195 \\
    $\hat{\sigma}_1$ & 1.3951 & 0.0209 & 1.3546 & 1.3947 & 1.4364 \\
    $\hat{\sigma}_2$ & 1.1700 & 0.0182 & 1.1352 & 1.1698 & 1.2059 \\
    $\hat{\sigma}_3$ & 0.3992 & 0.0008 & 0.3970 & 0.3994 & 0.4000 \\
    $\hat{\sigma}_{\beta}$ & 0.8900 & 0.1091 & 0.6982 & 0.8803 & 1.1283 \\
    $\hat{\sigma}_{\alpha}$ & 1.5378 & 1.1932 & 0.0950 & 1.3068 & 4.5089 \\
    $\hat{\rho}_{\gamma}$ & 0.5084 & 0.5363 & -0.7621 & 0.7546 & 0.9963 \\
    $\hat{\sigma}_{\gamma}$ & 0.0786 & 0.0591 & 0.0060 & 0.0673 & 0.2061 \\
    $\hat{\rho}_{\phi}$ & 0.4089 & 0.0714 & 0.2682 & 0.4113 & 0.5445 \\
    $\hat{\sigma}_{\phi}$ & 0.5260 & 0.0235 & 0.4801 & 0.5259 & 0.5714 \\
    $\hat{\rho}_{\delta}$ & 0.9712 & 0.0255 & 0.8992 & 0.9791 & 0.9963 \\
    $\hat{\sigma}_{\delta}$ & 0.2303 & 0.0356 & 0.1629 & 0.2297 & 0.3029 \\
    $\hat{\sigma}_{\psi}$ & 0.3726 & 0.0569 & 0.2498 & 0.3753 & 0.4774 \\
    $\hat{\rho}_{\omega}$ & 0.7106 & 0.0563 & 0.5961 & 0.7131 & 0.8104 \\
    $\hat{\sigma}_{\omega}$ & 0.3823 & 0.0299 & 0.3249 & 0.3819 & 0.4428 \\
    \hline \hline \hline
\end{tabular}
\caption{Posterior estimates of the parameters of the irregular data model (IDM) for the AFR region. \vspace{0.2cm}}
\label{tab:AFROparamest}
\end{table}

In Table \ref{tab:AFROparamest}, we report estimates of parameters of the model for AFR region. Parameter estimates for other regions are presented in supplementary Tables 1 - 5. We observe that estimates of the source-specific intercept terms, $\hat{\lambda}^{(a)}$ and $\hat{\lambda}^{(o)}$), are consistently positive while those of $\hat{\lambda}^{(s)}$ are consistently negative in all the regions. However, only $\hat{\lambda}^{(s)}$ was significant in some of the regions. This further demonstrates that survey data where available, on average, tend to have lower values than admin and official data. Estimates of the residual standard deviation for survey data are all close to the upper bound of the prior placed on $\sigma_3$, demonstrating the strong effect of the prior in the fitted models. However, $\hat{\sigma}_1$ and $\hat{\sigma}_2$ are considerably higher than $\hat{\sigma}_3$ in all the regions except AMR and EUR. In both regions, the estimates of these parameters are very close, indicating that survey estimates are, on average, more likely to have greater or similar variation as other data sources in both regions (see e.g., Figure \ref{fig:regboxplot}).

When considering the main effects - $\beta_i$, $\alpha_j$ and $\gamma_t$ - these results indicate that in the AFR region, the vaccine random effect, $\alpha_j$, accounted for much (74.8\%) of the total variation $(\hat{\sigma}^2_\beta + \hat{\sigma}^2_\alpha + \hat{\sigma}^2_\gamma)$ explained by these terms. For other regions, the vaccine random effect accounted for between 79.1\% and 99.6\% of the total variation explained by the main effects. This demonstrates substantial variation in coverage levels between the vaccines across all the regions. There is also considerable variation in coverage levels among countries within each region, as the estimates of $\sigma_\beta$ show. However, the temporal main effect, $\gamma_t$, explains very little variation in the data in each region (except for the AFR region), which is likely due to the significant effect of temporally correlated interaction terms in the models. Similarly, when considering the estimated variances of the interaction terms, the country-time interaction, $\phi_{it}$, explained the most variation in the data compared to other interaction terms in the AFR region. This was also the case in the SEAR region. For AMR, EMR and WPR regions, the most variation was explained by the country-vaccine-time interaction $\omega_{ijt}$, whereas for the EUR region, this was explained by the country-vaccine interaction, $\psi_{ij}$. In all the regions, the vaccine-time interaction, $\delta_{jt}$, explained the least variation in the data compared to other interaction terms. These results indicate the presence of strong within-country trends in the data. Estimates of the autoregressive parameter for $\gamma_t$, $\hat{\rho}_\gamma$, are not significant across all the regions, further indicating the little contribution of this global term in the fitted models (although its inclusion supports the structure of the interaction terms). However, the autocorrelation parameters of all the time-varying interaction terms are estimated to be significantly positive in all the regions, suggesting strong positive temporal trends which are tied to other sources of variation (country and vaccine) in the data.

\subsection{Modelled estimates of national immunization coverage and comparisons with WUENIC estimates}

\begin{figure}[ht]
    \centering
 \includegraphics[width=1.0 \textwidth]{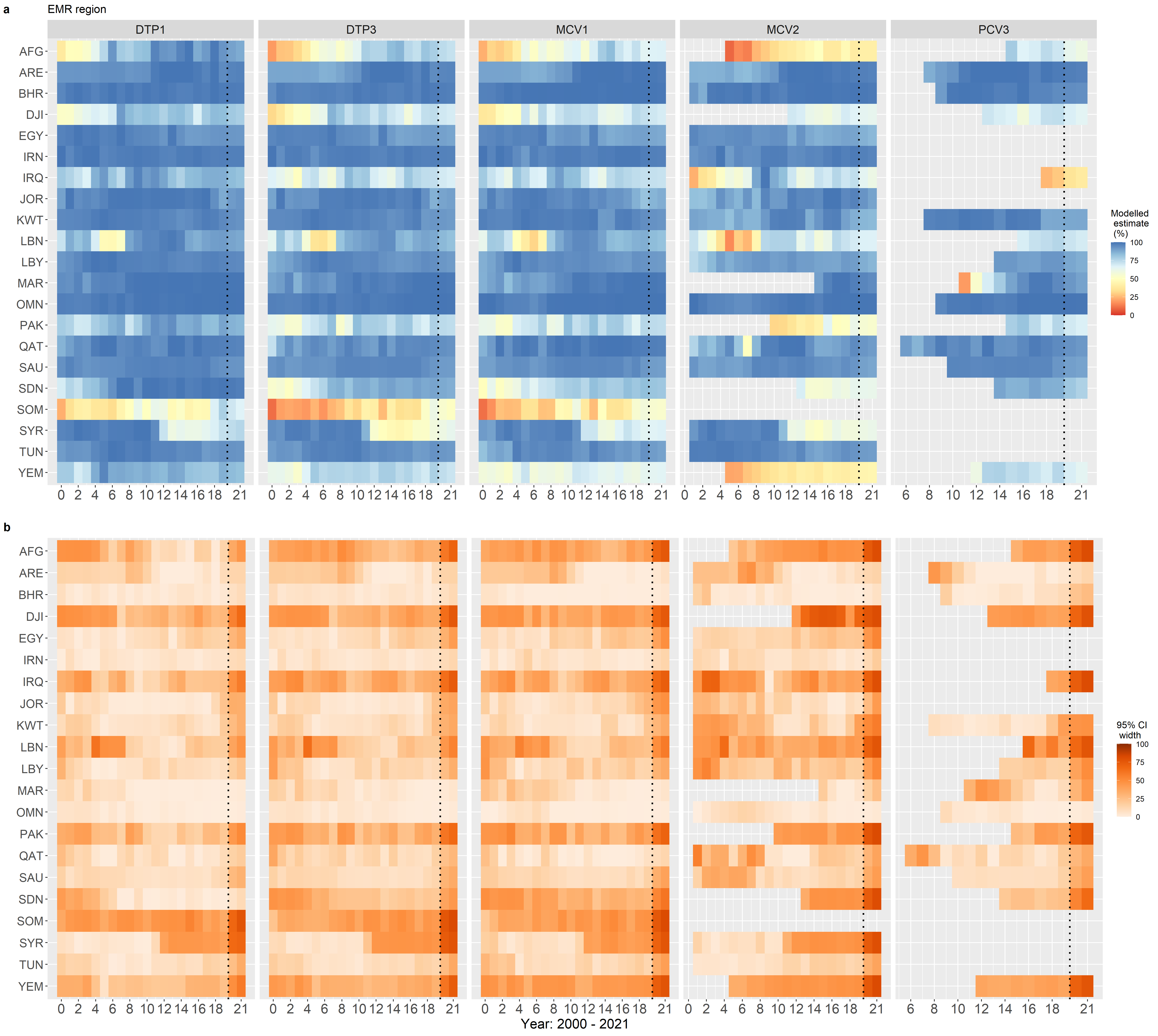} 
 \caption{ Modelled estimates of immunization coverage (a) and corresponding uncertainty estimates (b) for the EMR region. Predictions for 2021 and 2022 are shown on the right-hand side of the black dotted vertical lines.
\label{fig:emroctry}} 
\end{figure}

In Figure \ref{fig:emroctry} and supplementary Figures 6 - 10, we present plots of modelled estimates of coverage and associated uncertainties for EMR and other WHO regions, respectively. Time series plots of the modelled estimates overlaid with the input admin, official and survey data, as well as corresponding WUENIC estimates (2020 revision published in 2021) are also shown in supplementary Figure 11. In general, we observe that the patterns in immunization coverage are similar for DTP1, DTP3 and MCV1 due to these vaccines being introduced much earlier (since the 1970s) in the study countries, and therefore exhibit more stable trends, compared to MCV2 and PCV3. Both newer vaccines tend to have different patterns since these are mostly driven by the length of time since introduction and the speed of uptake. We also note that the fitted models produced plausible estimates of coverage relative to the input data and adjust well to survey data in some cases for the example countries as shown in supplementary Figure 11. For EMR, coverage appears high and more stable in many countries, although some countries had lower coverage at the beginning of the study period (e.g. AFG and DJI), but made substantial progress which appears to stagnate in latter years. The uncertainty estimates are low in most cases but generally higher for country-vaccine-time combinations for which input data were scanty or unavailable (e.g., DJI-MCV2 in Figure \ref{fig:emroctry}) and for out-of-sample predictions. In general, the predictions for 2020 and 2021 show changes in coverage, but not substantially from the preceding years. We note that these predictions did not take into account any disruptions to routine immunization caused by the pandemic. Hence, they represent a counterfactual non-pandemic scenario.

\begin{figure}[htp]
\centering
\begin{tabular}{l}
  \includegraphics[width=0.8 \textwidth]{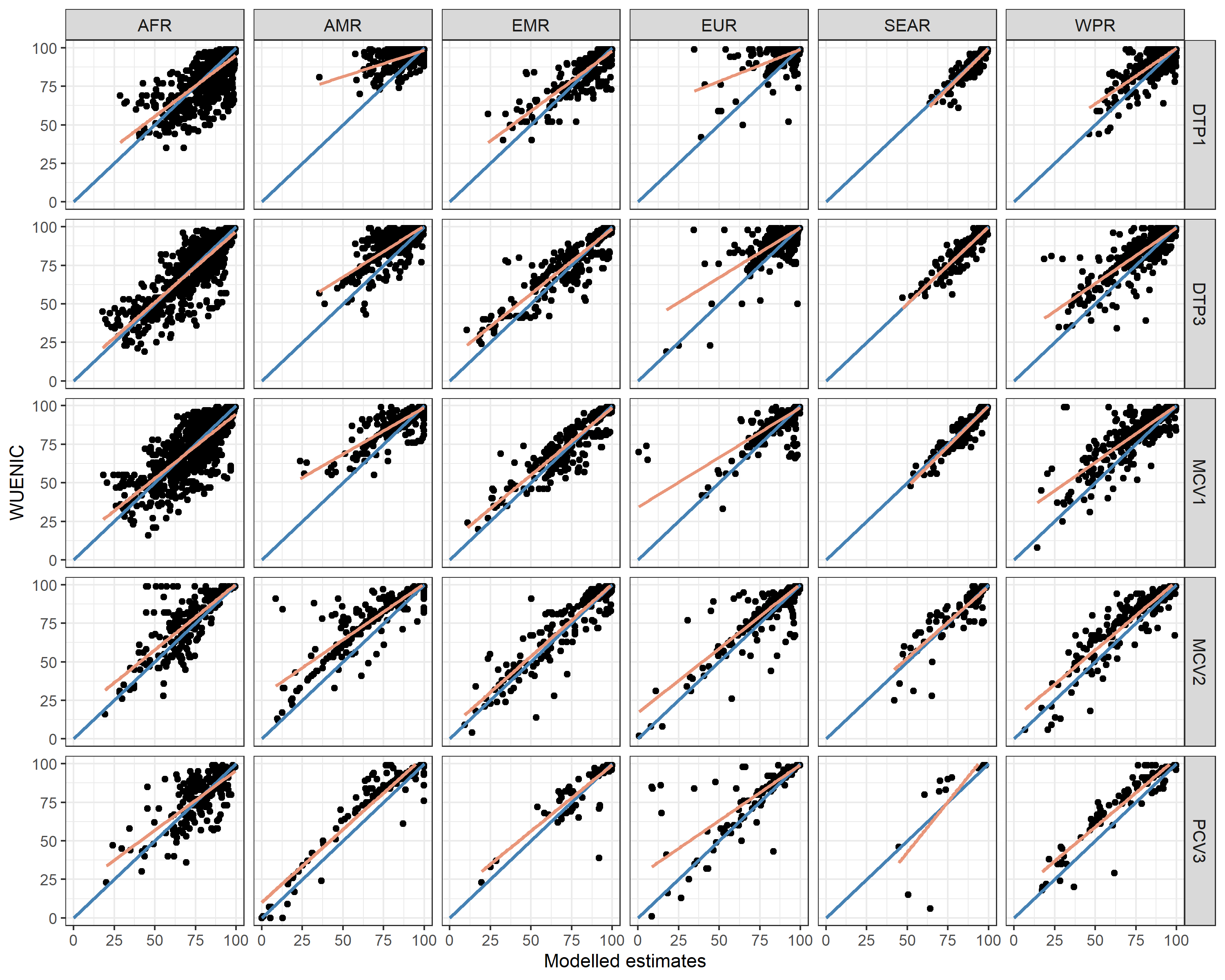} 
\end{tabular}
\caption{Plots of WUENIC versus modelled estimates by WHO regions and vaccine type. The blue lines illustrate a perfect agreement between both estimates while the light red lines are simple least square fits to the estimates within each region.} 
\label{Fig:wueniccomp2} 
\end{figure}

In Figure \ref{Fig:wueniccomp2} and supplementary Figures 12 - 18, we show comparisons between WUENIC and the modelled estimates for the period 2000 to 2019. There is generally a good level of correspondence between both estimates across the regions despite the differences in the methodologies used to produce these. The best correlations seem to have occurred in AFR, EMR and SEAR regions, but the most differences also occurred in AFR where the modelled estimates tend to be higher than WUENIC in some countries for DTP1, DTP3 and MCV1. At the global level, we obtained a median difference of -1.90\% with an interquartile range of 5.44\% (supplementary Table 6) between WUENIC and the modelled estimates. The highest median difference (in absolute value) was observed in AMR (-4.62\%) whilst the largest interquartile range was observed in AFR (10.35\%). Overall, these results strongly indicate that the modelled estimates are close to WUENIC.

Furthermore, the trends in the regional estimates of immunization coverage are displayed in supplementary Figures 19 (a) and (b). For the AFR region, for example, all five vaccines except MCV2 showed increasing trends which appear to level off or regress towards the end of the study period. Similar or different patterns were observed in other regions. The uncertainties associated with these regional estimates show robust estimation in most cases. 

\section{Software}
\label{sec:software}
We developed the {\tt imcover} package in the R statistical programming language to support the proposed Bayesian modeling approaches for immunization coverage. {\tt imcover} is an interface to the Stan programming language implementing the No-U-Turn Sampler (NUTS). The package includes functionality to replicate the analyses described in \ref{sec:method}, including both the BDSL and IDML models. The package is designed to support reanalysis of the WUENIC data with functionality to retrieve input data from the WHO website, process the files, fit models and produce coverage estimates for required country- and regional-vaccine-time combinations. The software is available from \url{https://wpgp.github.io/imcover/} and is described in more detail along with a full processing script in Supplemental materials (Sections 2 and 4).

\section{Discussion}


We have laid out a new methodology for producing ENIC, as an alternative or a complement to the WUENIC approach. Our methodology is based on a Bayesian hierarchical model which accounts for the full range of uncertainties present in the input data as well as those associated with the modelled estimates. The methodology is implemented in a user-friendly R package {\tt imcover}.  We investigated two candidate models and concluded that the irregular data multiple likelihood (IDML) model performed better both in terms of model fit and predictive performance. Regarding computing time, the IDML model takes an average of 1.6 hours to run 4,000 iterations, which includes a burn-in of 2,000 iterations, on a high specification computer, for each WHO region.

The work presented here is an improvement over previous work \citep{UtaziWHO2020}, which utilized some of the rules implemented in the WUENIC computational logic approach (e.g., denominator adjustment for admin estimates and choosing survey data when the differences between survey estimates and denominator-adjusted admin estimates were $> 10\%$) to process and harmonize the multiple input data. The processed data was then modelled using a BHM similar to the models developed in the current work, but only accounting for (random) spatial, temporal and vaccine-related variations and their interactions, to obtain smoothed coverage estimates and associated uncertainties. Whilst this approach produced coverage estimates that were similar to WUENIC, the ad hoc method adopted in combining information from the multiple input data prior to model-fitting meant that the full range of variability in the data was not properly accounted for. Our methodology also offers significant improvements over the WUENIC approach. It provides a mechanism to: estimate the uncertainties associated with the modelled estimates; borrow strength across countries, vaccine and time to improve coverage estimation; and predict immunization coverage for future time points. Also, additional informed beliefs could be introduced in a more methodical manner in the modelling stage through prior specifications on the parameters of the model. Another model-based approach for producing ENIC was developed by \cite{Limetal2008}, but this was based predominantly on survey data and was implemented in a non-Bayesian framework which does not allow the incorporation of prior beliefs into the modelling process. Also, \cite{GBDVax2021} utilized data from official sources and surveys to produce ENIC using a model-based approach, but their methodology was based on fitting spatio-temporal Gaussian process regression  models in multiple steps: first, for bias-adjustment of official data using survey dat, and then joint modelling of bias-adjusted official and survey data.

The quality of the modelled estimates produced in our work is largely dependent upon the quality and degree of completeness of the input data. First, as we noted in Section \ref {sec:data}, all three input data sets were not simultaneously available for desired country-time-vaccine combinations, with survey estimates being the most incomplete data source and unequally distributed over time and by WHO region. There were also instances where no data were available from all three data sources. As we observed in Section \ref{sec:application}, the proposed methodology is more likely to produce more robust and more precise estimates when more (accurate) input data are used for model-fitting. Secondly, the different input data sets have their inherent biases (e.g., admin estimates being greater than 100) which are likely the result of inaccurate denominator and/or numerator estimates, large differences between consecutive coverage estimates (in time)  and recall bias associated with survey data for multi-dose vaccines \citep{Cuttsetal2016}. A complete overview and analysis of data quality issues associated with these data sources are provided in \cite{Rauetal2022,Stashkoetal2019}. Although we implemented some ad hoc measures to correct these biases wherever possible, e.g., recall-bias adjustment for survey estimates of DTP3 and PCV3 and rounding down of administrative estimates greater than 100 whilst persevering the differential between multi-dose vaccines, they are better addressed at the point of data collection and summarization. Hence, efforts should be intensified within countries to improve the quality of data collected via these sources as has been recommended by global advisory bodies \citep{SCOBIEetal2020}. 

Wherever possible, the data processing steps presented here could be much improved to deal with any remaining data quality issues prior to model-fitting, as we have highlighted previously. It is possible to `switch-off' an entire data source for a given country-vaccine combination if it is deemed unfit for model-fitting, and then utilize input data from other sources for coverage estimation. Making such decisions on a case-by-case basis, though arduous, could lead to better quality modelled estimates that reflect the peculiarities of individual countries. There is, however, a need to strike a balance between how much data are available for model-fitting and the quality of modelled estimates that is desired. Additional data processing steps could also include expertly identifying and excluding any implausible outlying observations (e.g., some coverage estimates $\approx 1\%$ included in the current analysis) that could bias the modelled estimates. 


The modelling approaches outlined here are subject to other limitations. Currently, our methodology does not utilize any covariate information for coverage estimation. The inclusion of highly informative covariates such as access to a health facility, antenatal care attendance and female literacy rates \citep[see, e.g.,][]{Utazietal2022, GBDVax2021} in the models could help improve the accuracy and precision of the estimates. This will be fairly straightforward to implement, but we anticipate that it will greatly simplify the structures of some of the terms used to account for residual variation in models (\ref{eq:single1}) and (\ref{eq:eq1}), or make these redundant. An effective model selection criteria will, therefore, be needed to determine the best model parameterization in this setting. Furthermore, our modelled estimates had wide uncertainties in some cases, particularly where input data were scarce. The amount of uncertainty present in the modelled estimates could be further controlled by adjusting the priors on $\lambda^{(a)}$, $\lambda^{(o)}$ and $\lambda^{(s)}$ in the IDML model. 

Future work will focus on extending the proposed methodology to subnational coverage estimation, which is highly relevant to current global health agenda \citep{SDG2015,IA2030}. Whilst this will present additional challenges \citep[see, e.g.,][]{Brown2018}, we anticipate that subnational data will have a richer spatial structure unlike country-level data, which can be accounted for using conditional autoregressive priors \citep{Lee2011}. Furthermore, our predictions for 2020 and 2021 - which represent a counterfactual non-pandemic scenario - can be used to evaluate the impact of the Covid-19 pandemic on routine immunization coverage. Lastly, we will extend our analyses to include other vaccines not used for model development. 

In conclusion, our work holds a lot of promise as it adds to increased efforts to boost the quality of immunization coverage estimates available for global health policy and decision-making.

\section*{Acknowledgements}
This work was funded by WHO (Grant numbers 2020/1077452-0 and 2021/1103498-0), and in part by the Bill and Melinda Gates Foundation (Grant number INV-003287), and carried out at WorldPop, University of Southampton, United Kingdom. The authors gratefully acknowledge the WHO-UNICEF immunization coverage working group for their valuable inputs and feedback during model and software development. The authors also acknowledge the use of the IRIDIS High Performance Computing Facility, and associated support services at the University of Southampton, in the completion of this work.

\bibliographystyle{rss} 
\bibliography{REFERENCES}
\end{document}


\quad \\
\quad \\
\noindent This document accompanies the main paper. It contains additional information, including additional tables and figures referenced in the paper.

\section{Recall bias adjustment and processing of survey data}
For survey data, we implemented an additional data cleaning step to ensure consistency in the entries in each column, particularly where these were character variables. Each coverage survey estimate was then linked to a 'birth cohort year' which we used as the reference year for the estimate in this work. The birth cohort year was determined using the period of data collection and the age of the birth cohort that the survey estimate relates to, as in WUENIC methodology.

Similar to WUENIC approach, we applied a recall-bias adjustment to DTP3 and PCV3 survey estimates. Estimates based on vaccination cards only or vaccination cards and recall were used for the adjustment. In the pre-cleaned input data file, these estimates were labelled as: ``Card'' and ``Card or History'', respectively, in the column for evidence of vaccination. For country-vaccine-year combinations with multiple estimates labelled as ``crude'' or ``valid'', the ``valid'' estimates were retained in the analysis as these were considered more accurate. The formula used for the adjustment is:
\begin{equation}
    \textrm{VD3}_{\textrm{(card+history)}} = \textrm{VD3}_{\textrm{(card only)}} \times \frac{\textrm{VD1}_{\textrm{(card+history)}}}{\textrm{VD1}_{\textrm{(card only)}}}
\label{eq:eq1}
\end{equation}
where VD3 denotes the third dose of DTP or PCV vaccine and VD1, the first dose. We note that for each vaccine, the adjustment was applied only when all the data needed to compute equation (\ref{eq:eq1}) were available. After the adjustment, the original ``Card or History'' survey estimates of DTP3 and PCV3 were replaced with corresponding bias-adjusted estimates for further processing.

For a given vaccine, country and year, if one survey estimate was available, it was accepted if the sample size was greater than 300 or if the estimate was labelled 'valid'. Otherwise, the estimate was not accepted. Where multiple estimates were available for the same vaccine, country and year, ``Card or History'' estimates were prioritized over ``Card'' only estimates, and either of these were accepted if the corresponding sample size was greater than 300 or if the evidence of vaccination was based on valid doses. We note that for DTP3 and PCV3, only the bias-adjusted estimates were considered when available. When multiple estimates were available from the preceding step (perhaps from different surveys or the same survey) for the same vaccine, country and year, the estimate with the largest sample size was accepted. If the sample sizes were missing, the first valid estimate or the first estimate available was chosen, in the given sequence. The resulting survey estimates were used in the rest of the analyses.

\section{Software}
\label{sec:software}
In order to support the reproducibility and replication of the immunisation coverage modelling methods described in this report, we developed a set of tools in the R programming language \citep{Rlang2021}. The {\tt imcover} package provides functions for assembling the common sources of immunisation coverage and for fitting the blanced data single likelihood (BDSL) and irregular data multiple likelihood (IDML) models described in Section 3 using full Bayesian inference with Stan \citep{stan}. The latest version of imcover can be installed from GitHub by typing the command within the R console: \\ \texttt{devtools::install\_github(`wpgp/imcover')}.

In order to properly install the package, a C++ compiler is required. Internally, {\tt imcover} relies on Stan code which is translated into C++ and compiled, allowing for faster computations. On a Windows PC, the Rtools program provides the necessary compiler. This is available from \url{https://cran.r-project.org/bin/windows/Rtools/} for R version 4.0 or later. On Mac OS X, users should follow the instructions to install XCode. Users are advised to check the Stan installation guide for further information on the necessary system set-up and compilers (\url{https://github.com/stan-dev/rstan/wiki/RStan-Getting-Started}).

\begin{figure}[ht]
    \centering
 \includegraphics[width=1.0 \textwidth]{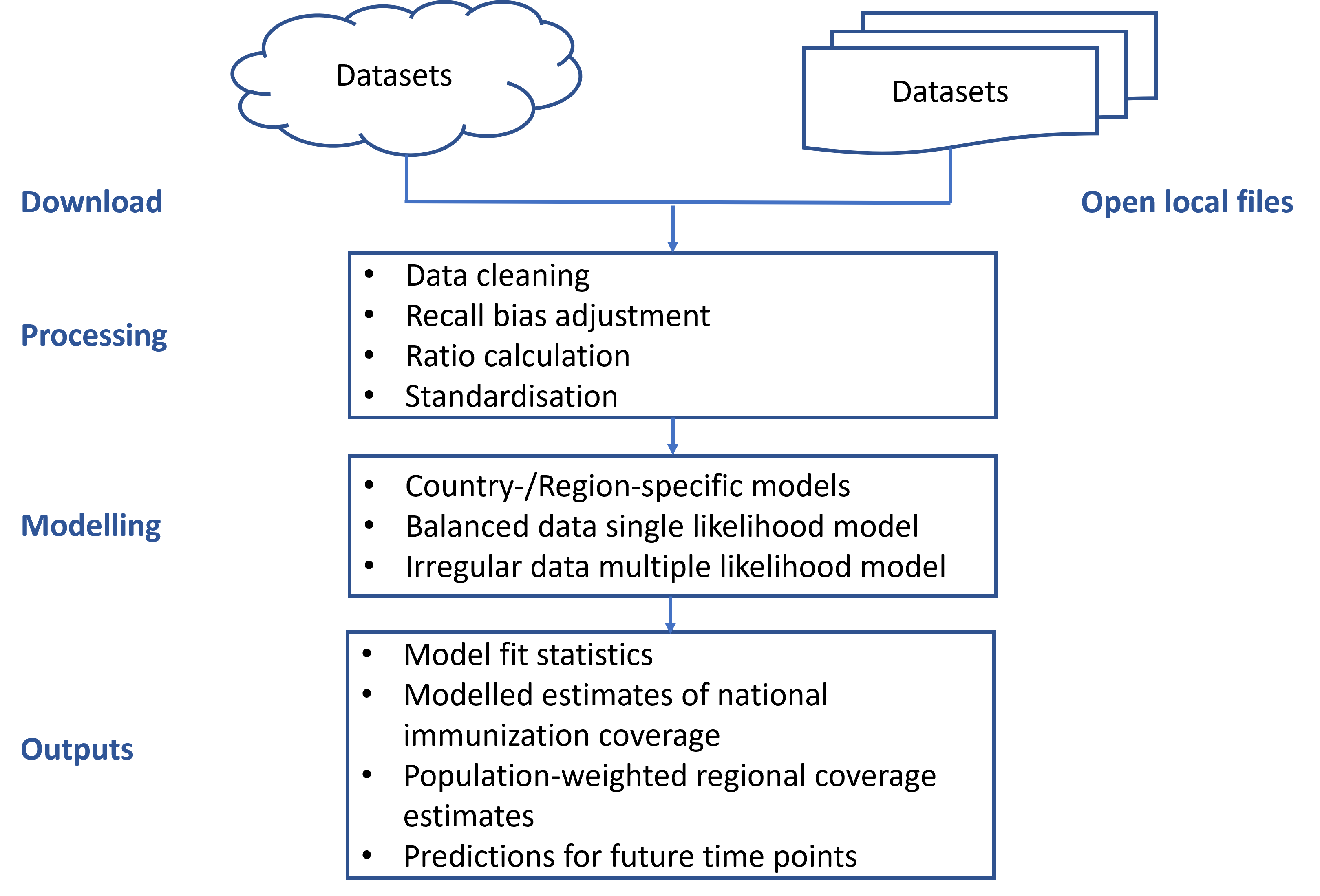} 
 \caption{Overview of steps supported by the {\tt imcover} package.
\label{fig:imcoveroverview}} 
\end{figure}

A typical workflow using imcover to produce national- and regional-level immunisation coverage estimates is illustrated in  Figure \ref{fig:imcoveroverview}. The user should first download the time series of reported coverage data, which may come from multiple sources (i.e. administrative, official, and survey estimates). Second, these datasets are processed in several steps to filter, clean duplicate records and correct possible reporting biases and then assemble a single dataset. This stage  creates datasets of format \texttt{ic.df} within R. This format is an extension of the common data frame and enables some of the specialised processing steps by {\tt imcover}. Third, the model is fitted against the assembled coverage dataset. The user has the option at this stage to specify additional parameters and prior choices for the statistical model. Fourth, after the model has been fitted, a model object is returned. This object's class extends the model objects from \texttt{rstan} from which parameter estimates and other results can be extracted in R. In the fifth stage, post-processing is done on the model results. These functions provide support to produce summaries of the estimates, predictions forward in time, standardised visualisations, and population-weighted regional aggregations of immunisation coverage estimates. The details of this workflow are illustrated below with a worked example of coverage data for the WHO AFRO region. Further information on the R package can be found within the documentation, including a long-form vignette, see $help(imcover)$.\\

\noindent {\bf Workflow example}\\
In the following sections we provide a worked example to produce time series of modelled estimates of immunisation coverage using the model-based approach implemented in R using {\tt imcover}. After loading the package, we first obtain the data from the WHO Immunisation Data Portal (\url{https://immunizationdata.who.int}). An internet connection is required for this step.\\

\begin{lstlisting}[language=R]
# load the package within the R environment
library(imcover)

# download administrative and official records
cov <- download_coverage()

# download survey records
svy <- download_survey()
\end{lstlisting}

Data downloading is handled by two functions: \texttt{download\_coverage} and \texttt{downloa d\_survey}. These handle administrative/official estimates or survey datasets, respectively. By default, the downloaded files are stored as temporary documents in the user's R temporary directory; however, the functions provide the option for users to save the downloaded files to a user-specified location and load them later from a local file path. In this way, a user can also load their own source of immunisation coverage data and process it into a standardised format for modeling.\\

\noindent {\bf Data processing and formatting}\\
As part of the download function, a series of checks and cleaning steps are applied by default. The goal of these checks is to identify the core attributes in the input data necessary to construct an immunisation coverage dataset. Specifically, these attributes include a country, time, vaccine identifier, and percent of the population covered by that vaccine. In the absence of a reported coverage percentage, the number of doses administered and target population can be used to estimate coverage. Identifying the core attributes allows the user to harmonise multiple source files into a unified data format for modelling. Administrative and official coverage estimates are processed similarly. Within \texttt{download\_survey}, the household survey datasets require some more specialised processing. For instance, multi-dose vaccine reports can have reporting and recall biases. Note that all the processing steps can be carried out using separate functions available in {\tt imcover} if advanced users want more control over pre-processing.

Within the {\tt imcover} package, processed coverage datasets are stored in \texttt{ic.df} format, or an ``immunisation coverage data frame". This format extends the common data frame object of R where observations are rows and attributes are stored in columns. \texttt{ic.df} objects support all standard methods and ways of working with data frames in R. This includes selecting records and columns by indices or column names, merging data frames, appending records, renaming, etc. The advantage of the \texttt{ic.df} format over a standard data frame is that it includes information to identify the columns containing core coverage information as well as notes on data pre-processing that has been done. These allow users to combine disparate sources of information on immunisation coverage into a harmonised analysis dataset without having to adjust for missing or differently named columns.

\begin{lstlisting}[language=R]
# note the type of object created
class(cov)
#> [1] "ic.df"   "data.frame"
\end{lstlisting}

The data files available from the WHO website require some additional cleaning before analysis. Notice that the data objects created by {\tt imcover} can work with all standard R commands.

\begin{lstlisting}[language=R]
# Further data cleaning of immunization records
# drop some record categories (PAB, HPV, WUENIC)
cov <- cov[cov$coverage_category %in% c("ADMIN", "OFFICIAL"), ]
cov$coverage_category <- tolower(cov$coverage_category)  # clean-up

# create a combined dataset
dat <- rbind(cov, svy)

# remove records with missing coverage values
dat <- dat[!is.na(dat$coverage), ]

# mismatch in vaccine names between coverage and survey datasets
dat[dat$antigen == 'DTPCV1', 'antigen'] <- 'DTP1'
dat[dat$antigen == 'DTPCV2', 'antigen'] <- 'DTP2'
dat[dat$antigen == 'DTPCV3', 'antigen'] <- 'DTP3'

# subset records
dat <- ic_filter(dat, 
                 vaccine = c("DTP1", "DTP3", "MCV1", "MCV2", "PCV3"),
                 time = 2000:2020)
\end{lstlisting}

In preparation for the statistical modelling we carry out several additional pre-processing steps. Firstly, some records observe inconsistencies in the levels of coverage between multi-dose vaccines. To maintain consistency, where coverage of later doses cannot exceed earlier doses, we model the ratio between first and third dose. In this example, we only adjust DTP1 and DTP3, but other multi-dose vaccines could be processed in a similar manner.

\begin{lstlisting}[language=R]
# adjustment - use ratio for DTP3
dat <- ic_ratio(dat, numerator = 'DTP3', denominator = 'DTP1')
\end{lstlisting}

The \texttt{ic.df} object will now store a note that this processing step has been
carried out so that the ratio is correctly back-transformed and that coverage estimates and predictions are adjusted appropriately.

Secondly, we need to force coverage estimates to lie between 0\% and 100\% so that we can model the data with a logit transformation.

\begin{lstlisting}
# maintain coverage between 0-100%
dat <- ic_adjust(dat, coverage_adj = TRUE)
\end{lstlisting}

\noindent {\bf Fitting models with {\tt imcover}}\\
The core of {\tt imcover} is the functionality to fit a Bayesian statistical model of multiple time series. A discussed in Section 3, this includes a BDSL and an IDML model. The sources of coverage data (in this example administrative, official and surveys) are taken as multiple, partial estimates of the true, unobserved immunization coverage in a country. A Bayesian estimation approach allows us to incorporate these multiple datasets, place prior beliefs on which sources are more reliable, share information between countries, and to quantify uncertainty in our estimate of the true immunization coverage.

{\tt imcover} provides an interface to Stan \cite{stan} for statistical computation. This means that, in addition to {\tt imcover}, many of the R programming language tools for assessing model performance and visualizing results from \texttt{rstan} will work for \texttt{imcover} results.

For this example we subset the records to the AFR region based on a regional identifier and the most recent years.

\begin{lstlisting}[language=R]
# enable parallel processing in rstan
options(mc.cores = parallel::detectCores())  

# Fit model to a single region of data
fit <- ic_fit(dat[dat$region == 'AFRO', ], 
              chains = 4, 
              iter = 2000, 
              warmup = 1000)
\end{lstlisting}

The previous command uses the IDML model, and  \texttt{ic\_fit\_single} can be used to fit a BDSL model to the same data. Other options to these fitting functions include, \texttt{prior\_sigma} which can be used to set the scale parameter for the truncated Cauchy priors on $sigma$ and \texttt{upper\_sigma} which allows users to set an upper bound on the scale parameter for each data source.

\noindent {\bf Model outputs}\\
After fitting the model, we can work with the parameter estimates and estimates of coverage. As noted previously, the fitted models returned by \texttt{ic\_fit} are \texttt{rstan}-class objects and are compatible with many other tools, such as \texttt{bayesplot}. {\tt imcover} wraps these objects into a new class, \texttt{ic.fit}, which, along with the fitted model object, contains additional information on the model type and data pre-processing steps to facilitate most common analyses of immunisation coverage.

For example, we can easily make graphs of coverage estimates using a generic \texttt{plot} function.

\begin{lstlisting}[language=R]
# Plot of coverage
plot(ic_filter(fit, vaccine = 'DTP1',
               country = c('BWA', 'SLE', 'GHA', 'NGA')),
    ncol = 2)
    
# Plot of coverage overlaid with WUENIC estimates
plot(ic_filter(fit, vaccine = 'DTP1',
               country = c('BWA', 'SLE', 'GHA', 'NGA')),
    ncol = 2,
    wuenic = TRUE)    
\end{lstlisting}

The main output of interest from {\tt imcover} is a table of coverage estimates for each country, vaccine, and time point, along with uncertainty around each estimate. We can extract these data as a table, save them to a  data frame, or write them out to a file for use in a report.

\begin{lstlisting}[language=R]
# Extracting coverage estimates to a data frame
ic <- ic_coverage(fit,
                  stat = "quantile", # customise the summary function
                  probs = c(0.025, 0.5, 0.975))
\end{lstlisting}

Population-weighted regional aggregations of immunisation coverage can also be calculated.

\begin{lstlisting}[language=R]
ic_regional(fit,
            stat = c("mean", "quantile"),
            probs = c(0.025, 0.5, 0.975)
\end{lstlisting}

One we have a fitted model for a specific time period, we can also use it to predict coverage to time point in the near future.

\begin{lstlisting}[language=R]
# Predict for future time points (3 years post 2018)
fit1 <- predict(fit1, t = 3) # update icfit to include 'prediction' info

# Update the graph to include fitted estimates, observed data, and predictions
plot(ic_filter(fit, vaccine = 'DTP1',
     country = c('BWA', 'SLE', 'GHA', 'NGA')),
     ncol = 2,
     predition = TRUE)
\end{lstlisting}

This section has introduced the core functionality of the {\tt imcover} package and its use in modelling national time series of immunization coverage. Providing a coordinated toolset to download, process, and model
immunization data should allow for consistent analyses whose methods are transparent and replicable.

\section{Supplementary tables and figures}
\begin{figure}[H]
    \includegraphics[width = \textwidth]{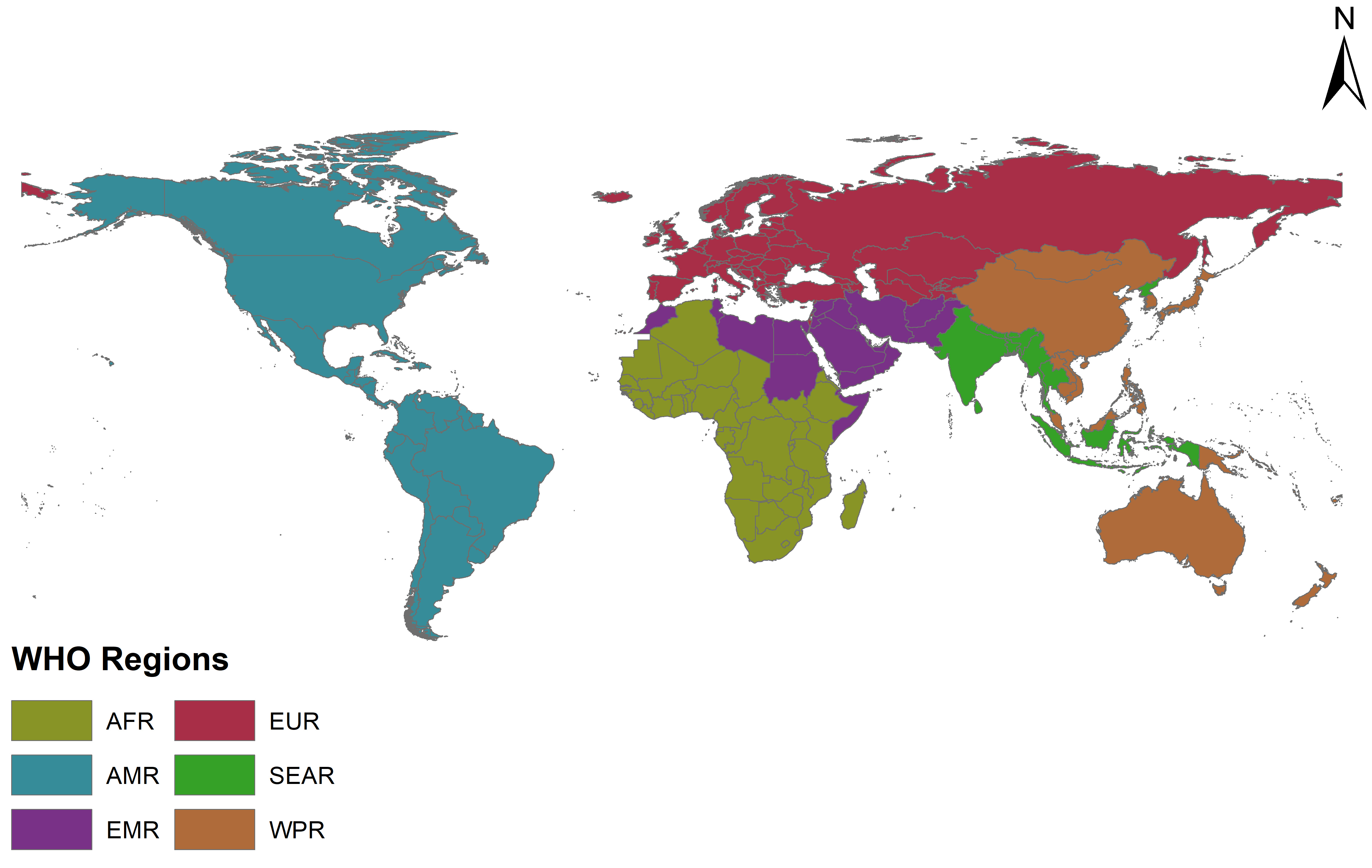}
    \caption{WHO Member States and regions.}
\end{figure}

\newpage
\begin{figure}[H]
    \includegraphics[width = \textwidth]{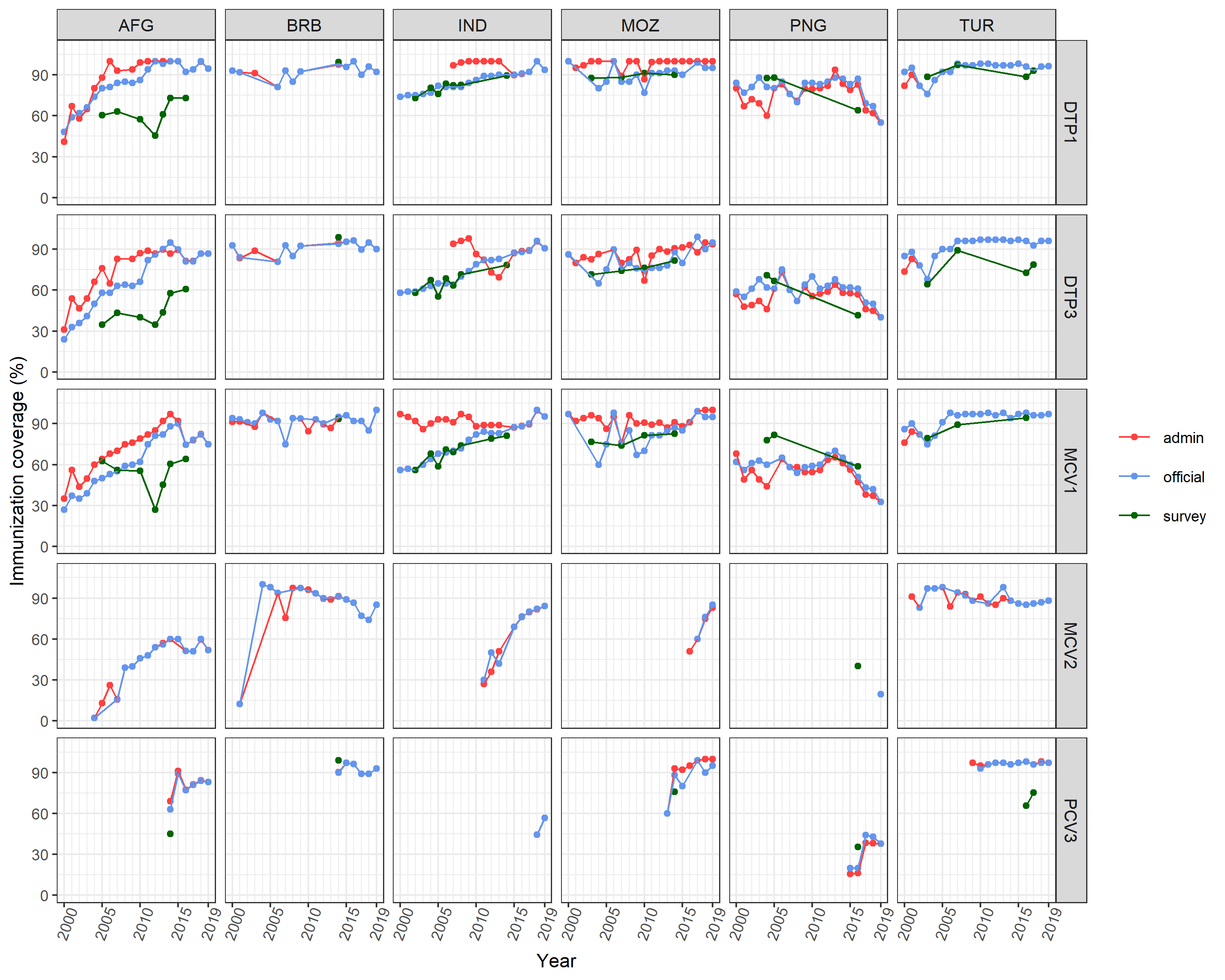}
    \caption{Time series plots of the processed input administrative, official and survey data for some example countries.}
\end{figure}

\newpage
\begin{figure}[H]
 \centering
 \includegraphics[width=1 \textwidth]{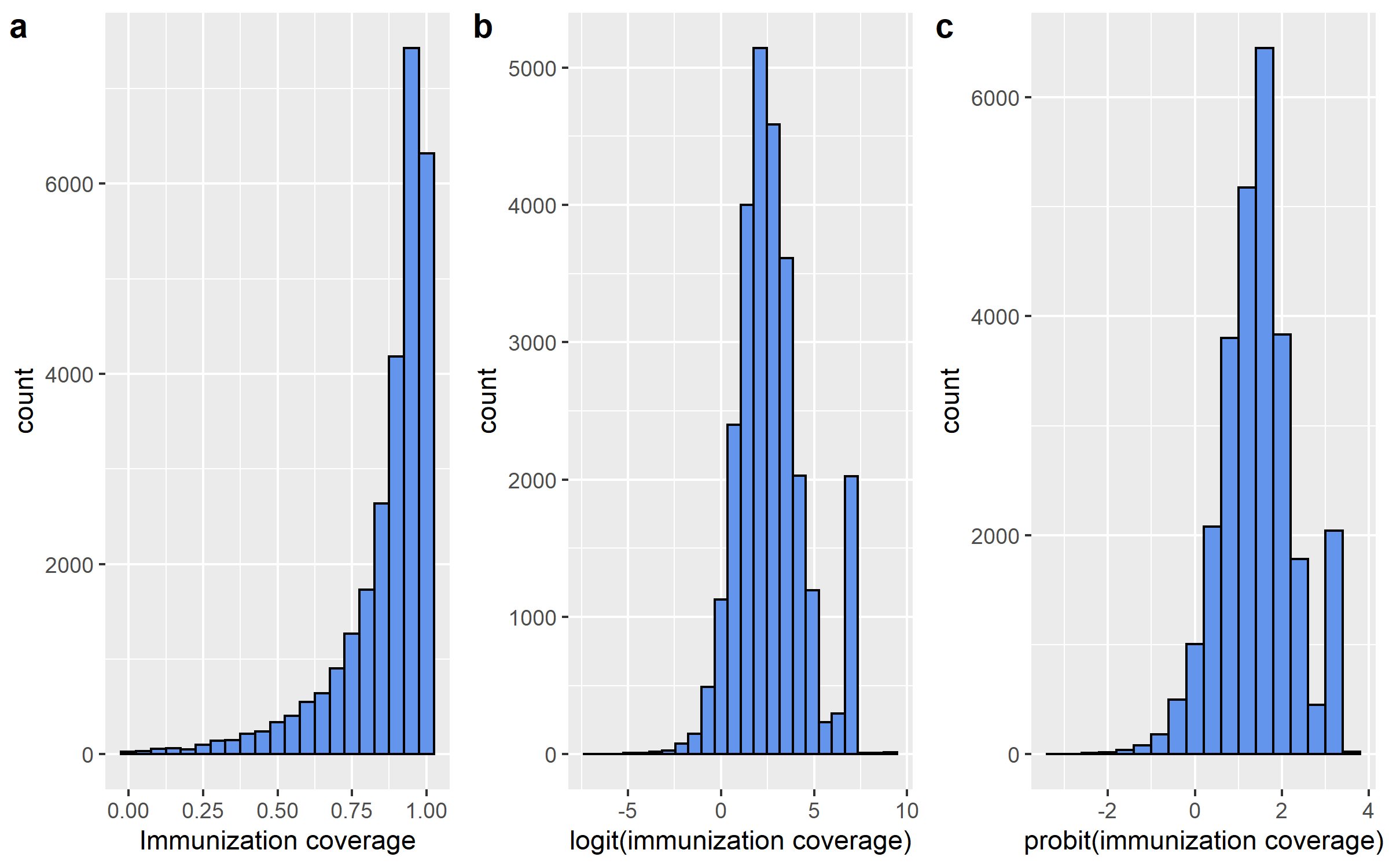} 
 \caption{Distributions of (a) original, (b) logit-transformed and (c) probit-transformed processed pooled national immunization coverage data.
\label{fig:histplot}} 
\end{figure}

\newpage
\begin{figure}[H]
    \centering
 \includegraphics[width=1.0 \textwidth]{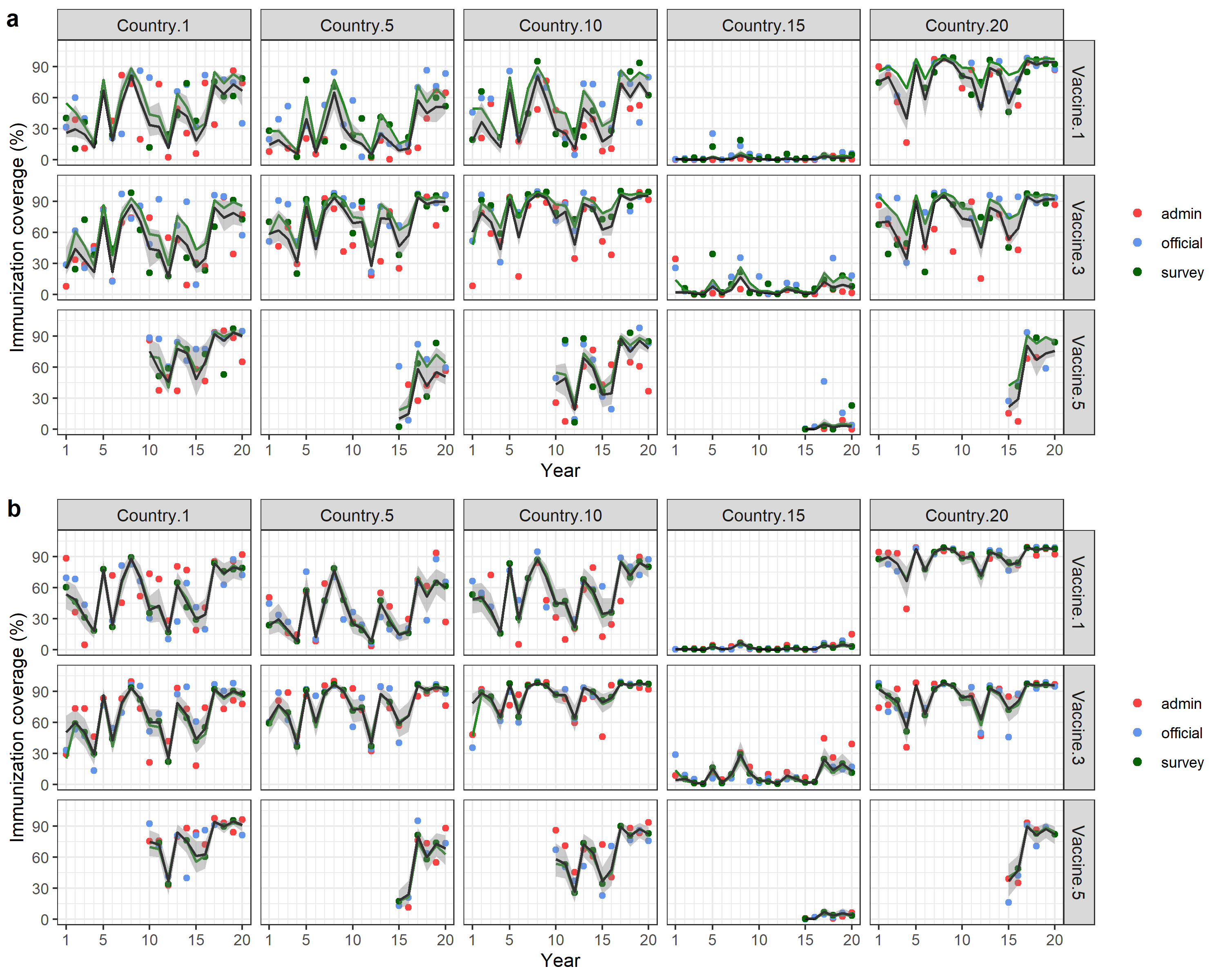} 
 \caption{Plots of simulated and modelled estimates of national immunization coverage for five countries and three vaccines using (a) the balanced data single likelihood (BDSL) model and (b) the irregular data multiple likelihood (IDML) model. The solid black lines and grey-shaded areas are the estimates of $p_{ijt}$ and the green lines are the true values. The data were generated under scenario 1 described in Section 5.
\label{fig:simplot1}} 
\end{figure}

\newpage
\begin{table}
\caption{Posterior estimates of the parameters of the multiple likelihood model for AMR region}
\centering
\begin{tabular}{ c  r  r  r  r  r }
    \hline \hline \hline
    \textbf{Parameter} & \textbf{Mean} & \textbf{Std. dev.} & \textbf{2.5\%} & \textbf{50\%} & \textbf{97.5\%} \\ [0.5 ex]
    \hline
    $\hat{\lambda}^{(a)}$ & 0.5008 & 2.858 & -0.0738 & 0.5090 & 1.0555 \\
    $\hat{\lambda}^{(o)}$ & 0.5017 & 0.2856 & -0.0741 & 0.5095 & 1.0574 \\
    $\hat{\lambda}^{(s)}$ & -0.6286 & 0.2854 & -1.2026 & -0.6203 & -0.0797 \\
    $\hat{\sigma}_1$ & 0.4215 & 0.0110 & 0.4011 & 0.4212 & 4439 \\
    $\hat{\sigma}_2$ & 0.4116 & 0.0109 & 0.3905 & 0.4118 & 0.4335 \\
    $\hat{\sigma}_3$ & 0.3998 & 0.0002 & 0.3993 & 0.3999 & 0.4000 \\
    $\hat{\sigma}_{\beta}$ & 0.4856 & 0.2770 & 0.0292 & 0.4972 & 1.0211 \\
    $\hat{\sigma}_{\alpha}$ & 3.0274 & 2.0690 & 0.5602 & 2.6560 & 7.7328 \\
    $\hat{\rho}_{\gamma}$ & 0.1471 & 0.5073 & -0.8011 & 0.1630 & 0.9543 \\
    $\hat{\sigma}_{\gamma}$ & 0.1066 & 0.0620 & 0.0057 & 0.1043 & 0.2381 \\
    $\hat{\rho}_{\phi}$ & 0.9502 & 0.0251 & 0.8882 & 0.9568 & 0.9808 \\
    $\hat{\sigma}_{\phi}$ & 0.2682 & 0.0329 & 0.2084 & 0.2670 & 0.3327 \\
    $\hat{\rho}_{\delta}$ & 0.9311 & 0.0645 & 0.7649 & 0.9501 & 0.9960 \\
    $\hat{\sigma}_{\delta}$ & 0.2696 & 0.0453 & 0.1873 & 0.2665 & 0.3559 \\
    $\hat{\sigma}_{\psi}$ & 0.7860 & 0.0687 & 0.6642 & 0.7833 & 0.9269 \\
    $\hat{\rho}_{\omega}$ & 0.3165 & 0.0249 & 0.2670 & 0.3169 & 0.3651 \\
    $\hat{\sigma}_{\omega}$ & 1.2770 & 0.0206 & 1.2367 & 1.2767 & 1.3152 \\
    \hline \hline \hline
\end{tabular}
\end{table}

\begin{table}
\caption{Posterior estimates of the parameters of the multiple likelihood model for EUR}
\centering
\begin{tabular}{ c  r  r  r  r  r }
    \hline \hline \hline
    \textbf{Parameter} & \textbf{Mean} & \textbf{Std. dev.} & \textbf{2.5\%} & \textbf{50\%} & \textbf{97.5\%} \\ [0.5 ex]
    \hline
    $\hat{\lambda}^{(a)}$ & 0.2615 & 0.2950 & -0.3199 & 0.2586 & 0.8420 \\
    $\hat{\lambda}^{(o)}$ & 0.2822 & 0.2951 & -0.3015 & 0.2792 & 0.8635 \\
    $\hat{\lambda}^{(s)}$ & -0.2474 & 0.2953 & -0.8318 & -0.2501 & 0.3380 \\
    $\hat{\sigma}_1$ & 0.4639 & 0.0102 & 0.4440 & 0.4637 & 0.4842 \\
    $\hat{\sigma}_2$ & 0.4639 & 0.0102 & 0.4440 & 0.4637 & 0.4842 \\
    $\hat{\sigma}_3$ & 0.3994 & 0.0006 & 0.3979 & 0.3996 & 0.4000 \\
    $\hat{\sigma}_{\beta}$ & 0.2410 & 0.1743 & 0.0105 & 0.2155 & 0.6291 \\
    $\hat{\sigma}_{\alpha}$ & 3.9051 & 2.2320 & 1.1716 & 3.4282 & 9.4420 \\
    $\hat{\rho}_{\gamma}$ & 0.1223 & 0.5833 & -0.8987 & 0.1599 & 0.9814 \\
    $\hat{\sigma}_{\gamma}$ & 0.0434 & 0.0282 & 0.0042 & 0.0383 & 0.1085 \\
    $\hat{\rho}_{\phi}$ & 0.9571 & 0.0114 & 0.9302 & 0.9585 & 0.9749 \\
    $\hat{\sigma}_{\phi}$ & 0.2472 & 0.0180 & 0.2125 & 0.2472 & 0.2823 \\
    $\hat{\rho}_{\delta}$ & 0.9769 & 0.0229 & 0.9141 & 0.9838 & 0.9991 \\
    $\hat{\sigma}_{\delta}$ & 0.1202 & 0.0207 & 0.0822 & 0.1193 & 0.1625\\
    $\hat{\sigma}_{\psi}$ & 0.7441 & 0.0613 & 0.6291 & 0.7428 & 0.8680\\
    $\hat{\rho}_{\omega}$ & 0.6527 & 0.0257 & 0.6032 & 0.6519 & 0.7037 \\
    $\hat{\sigma}_{\omega}$ & 0.6738 & 0.0125 & 0.6499 & 0.6736 & 0.6989 \\
    \hline \hline \hline
\end{tabular}
\end{table}

\begin{table}
\caption{Posterior estimates of the parameters of the multiple likelihood model for EMR}
\centering
\begin{tabular}{ c  r  r  r  r  r }
    \hline \hline \hline
    \textbf{Parameter} & \textbf{Mean} & \textbf{Std. dev.} & \textbf{2.5\%} & \textbf{50\%} & \textbf{97.5\%} \\ [0.5 ex]
    \hline
    $\hat{\lambda}^{(a)}$ & 0.5188 & 0.3051 & -0.0849 & 0.5231 & 1.1021 \\
    $\hat{\lambda}^{(o)}$ & 0.5189 & 0.3050 & -0.0774 & 0.5219 & 1.1056 \\
    $\hat{\lambda}^{(s)}$ & -0.6581 & 0.3054 & -1.2607 & -0.6528 & -0.0769 \\
    $\hat{\sigma}_1$ & 0.7680 & 0.0199 & 0.7298 & 0.7679 & 0.8077 \\
    $\hat{\sigma}_2$ & 0.6896 & 0.0189 & 0.6540 & 0.6893 & 0.7275 \\
    $\hat{\sigma}_3$ & 0.3990 & 0.0010 & 0.3963 & 0.3993 & 0.4000 \\
    $\hat{\sigma}_{\beta}$ & 1.3293 & 0.4876 & 0.1831 & 1.3620 & 2.2276 \\
    $\hat{\sigma}_{\alpha}$ & 2.5990 & 1.8366 & 0.6515 & 2.2021 & 7.0152 \\
    $\hat{\rho}_{\gamma}$ & 0.6071 & 0.4976 & -0.6624 & 0.8617 & 0.9990 \\
    $\hat{\sigma}_{\gamma}$ & 0.1149 & 0.0621 & 0.0153 & 0.1118 & 0.2426 \\
    $\hat{\rho}_{\phi}$ & 0.8985 & 0.0478 & 0.7949 & 0.9027 & 0.9717 \\
    $\hat{\sigma}_{\phi}$ & 0.4852 & 0.0375 & 0.4127 & 0.4846 & 0.5598 \\
    $\hat{\rho}_{\delta}$ & 0.3475 & 0.5466 & -0.8263 & 0.4814 & 0.9864 \\
    $\hat{\sigma}_{\delta}$ & 0.0540 & 0.0385 & 0.0042 & 0.0484 & 0.1379 \\
    $\hat{\sigma}_{\psi}$ & 0.5925 & 0.0769 & 0.4518 & 0.5881 & 0.7563 \\
    $\hat{\rho}_{\omega}$ & 0.4987 & 0.0458 & 0.4068 & 0.4998 & 0.5872 \\
    $\hat{\sigma}_{\omega}$ & 0.7634 & 0.0249 & 0.7148 & 0.7634 & 0.8131 \\
    \hline \hline \hline
\end{tabular}
\end{table}

\begin{table}
\caption{Posterior estimates of the parameters of the multiple likelihood model for SEAR}
\centering
\begin{tabular}{ c  r  r  r  r  r }
    \hline \hline \hline
    \textbf{Parameter} & \textbf{Mean} & \textbf{Std. dev.} & \textbf{2.5\%} & \textbf{50\%} & \textbf{97.5\%} \\
    \hline
    $\hat{\lambda}^{(a)}$ & 0.4308 & 0.2925 & -0.1347 & 0.4283 & 1.0051 \\
    $\hat{\lambda}^{(o)}$ & 0.1046 & 0.2916 & -0.4622 & 0.1020 & 0.6730 \\
    $\hat{\lambda}^{(s)}$ & -0.2016 & 0.2930 & -0.7794 & -0.2024 & 0.3754 \\
    $\hat{\sigma}_1$ & 1.4097 & 0.0417 & 1.3302 & 1.4091 & 1.4937 \\
    $\hat{\sigma}_2$ & 0.9489 & 0.0310 & 0.8900 & 0.9483 & 1.0120 \\
    $\hat{\sigma}_3$ & 0.3975 & 0.0025 & 0.3905 & 0.3982 & 0.3999 \\
    $\hat{\sigma}_{\beta}$ & 1.2842 & 0.4016 & 0.7332 & 1.2084 & 2.2674 \\
    $\hat{\sigma}_{\alpha}$ & 2.5514 & 1.6480 & 0.8068 & 2.1492 & 6.6952 \\
    $\hat{\rho}_{\gamma}$ & 0.8774 & 0.2489 & -0.0015 & 0.9595 & 0.9989 \\
    $\hat{\sigma}_{\gamma}$ & 0.1622 & 0.0606 & 0.0462 & 0.1581 & 0.2950 \\
    $\hat{\rho}_{\phi}$ & 0.4686 & 0.1939 & 0.0796 & 0.4738 & 0.8472 \\
    $\hat{\sigma}_{\phi}$ & 0.4585 & 0.0418 & 0.3786 & 0.4579 & 0.5420 \\
    $\hat{\rho}_{\delta}$ & 0.2717 & 0.3399 & -0.3255 & 0.2403 & 0.9580 \\
    $\hat{\sigma}_{\delta}$ & 0.1897 & 0.0485 & 0.0979 & 0.1882 & 0.2878 \\
    $\hat{\sigma}_{\psi}$ & 0.2524 & 0.1558 & 0.0167 & 0.2475 & 0.5612 \\
    $\hat{\rho}_{\omega}$ & 0.8599 & 0.0519 & 0.7290 & 0.8694 & 0.9354 \\
    $\hat{\sigma}_{\omega}$ & 0.3221 & 0.0393 & 0.2471 & 0.3208 & 0.4052 \\
    \hline \hline \hline
\end{tabular}
\end{table}

\begin{table}
\caption{Posterior estimates of the parameters of the multiple likelihood model for the WPR}
\centering
\begin{tabular}{ c  r  r  r  r  r }
    \hline \hline \hline
    \textbf{Parameter} & \textbf{Mean} & \textbf{Std. dev.} & \textbf{2.5\%} & \textbf{50\%} & \textbf{97.5\%} \\
    \hline
    $\hat{\lambda}^{(a)}$ & 0.5084 & 0.2952 & -0.0589 & 0.5050 & 1.0857 \\
    $\hat{\lambda}^{(o)}$ & 0.5584 & 0.2953 & -0.0062 & 0.5552 & 1.1388 \\
    $\hat{\lambda}^{(s)}$ & -0.6892 & 0.2961 & -1.2604 & -0.6901 & -0.1178 \\
    $\hat{\sigma}_1$ & 0.8704 & 0.0218 & 0.8273 & 0.8700 & 0.9135 \\
    $\hat{\sigma}_2$ & 0.8575 & 0.0212 & 0.8167 & 0.8574 & 0.8994 \\
    $\hat{\sigma}_3$ & 0.3989 & 0.0011 & 0.3960 & 0.3992 & 0.4000 \\
    $\hat{\sigma}_{\beta}$ & 1.2773 & 0.2275 & 0.8910 & 1.2561 & 1.7798 \\
    $\hat{\sigma}_{\alpha}$ & 3.1161 & 1.8398 & 0.7977 & 2.7570 & 7.6872 \\
    $\hat{\rho}_{\gamma}$ & 0.4081 & 0.5328 & -0.7494 & 0.5413 & 0.9989 \\
    $\hat{\sigma}_{\gamma}$ & 0.0722 & 0.0517 & 0.0051 & 0.0634 & 0.1939 \\
    $\hat{\rho}_{\phi}$ & 0.6871 & 0.0751 & 0.5332 & 0.6889 & 0.8303 \\
    $\hat{\sigma}_{\phi}$ & 0.5605 & 0.0377 & 0.4881 & 0.5607 & 0.6351 \\
    $\hat{\rho}_{\delta}$ & 0.5274 & 0.5582 & -0.7790 & 0.8258 & 0.9974 \\
    $\hat{\sigma}_{\delta}$ & 0.0913 & 0.0511 & 0.0021 & 0.0947 & 0.1878 \\
    $\hat{\sigma}_{\psi}$ & 0.5964 & 0.0868 & 0.4250 & 0.5956 & 0.7719 \\
    $\hat{\rho}_{\omega}$ & 0.6172 & 0.0410 & 0.5363 & 0.6168 & 0.6965 \\
    $\hat{\sigma}_{\omega}$ & 0.8232 & 0.0299 & 0.7642 & 0.8238 & 0.8820 \\
    \hline \hline \hline
\end{tabular}
\end{table}

\begin{table}[ht]
\caption{Summary of the differences between WUENIC and modelled estimates for each WHO region and globally. \vspace{0.2cm}}
\centering
\begin{tabular}{l r r} \hline \hline \hline
 Region & Median     & Interquartile \\
        & difference & range\\ \hline
 EMR   & -1.33 & 4.59 \\ 
 AMR   & -4.62 & 6.59 \\
 EUR   & -1.89 & 2.75 \\
 AFR   & 0.21 & 10.35 \\
 WPR   & -3.22 & 7.13 \\
 SEAR  & -0.36 & 3.22 \\ 
Global  & -1.90 & 5.44 \\ \hline \hline \hline
\end{tabular}
\label{tab:wueniccomptab}
\end{table}

\clearpage
\newpage
\begin{figure}[H]
    \includegraphics[width = \textwidth]{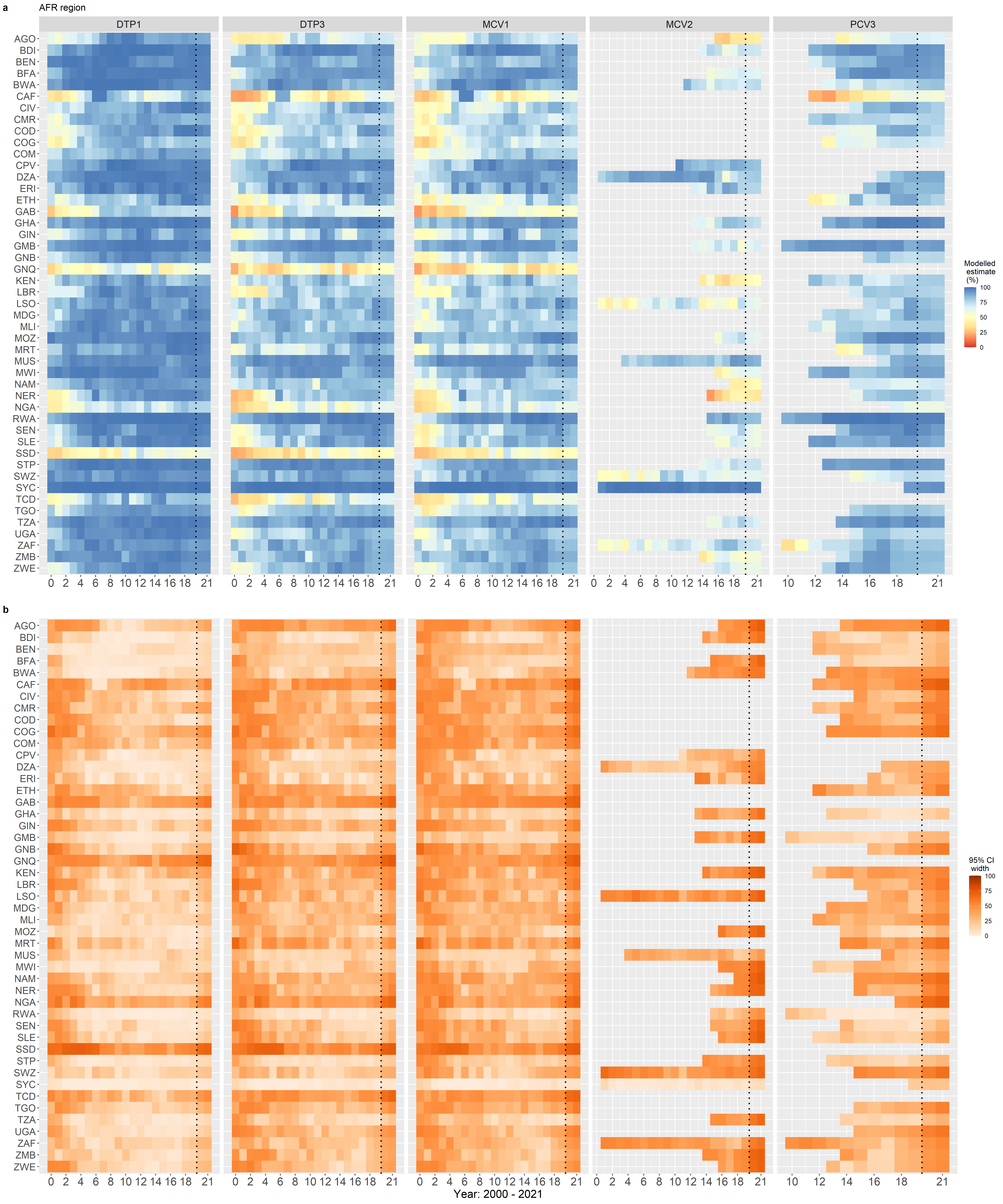}
    \caption{Plots of modelled estimates and corresponding uncertainties at the country level for AFR}
\end{figure}

\begin{figure}[H]
    \includegraphics[width = \textwidth]{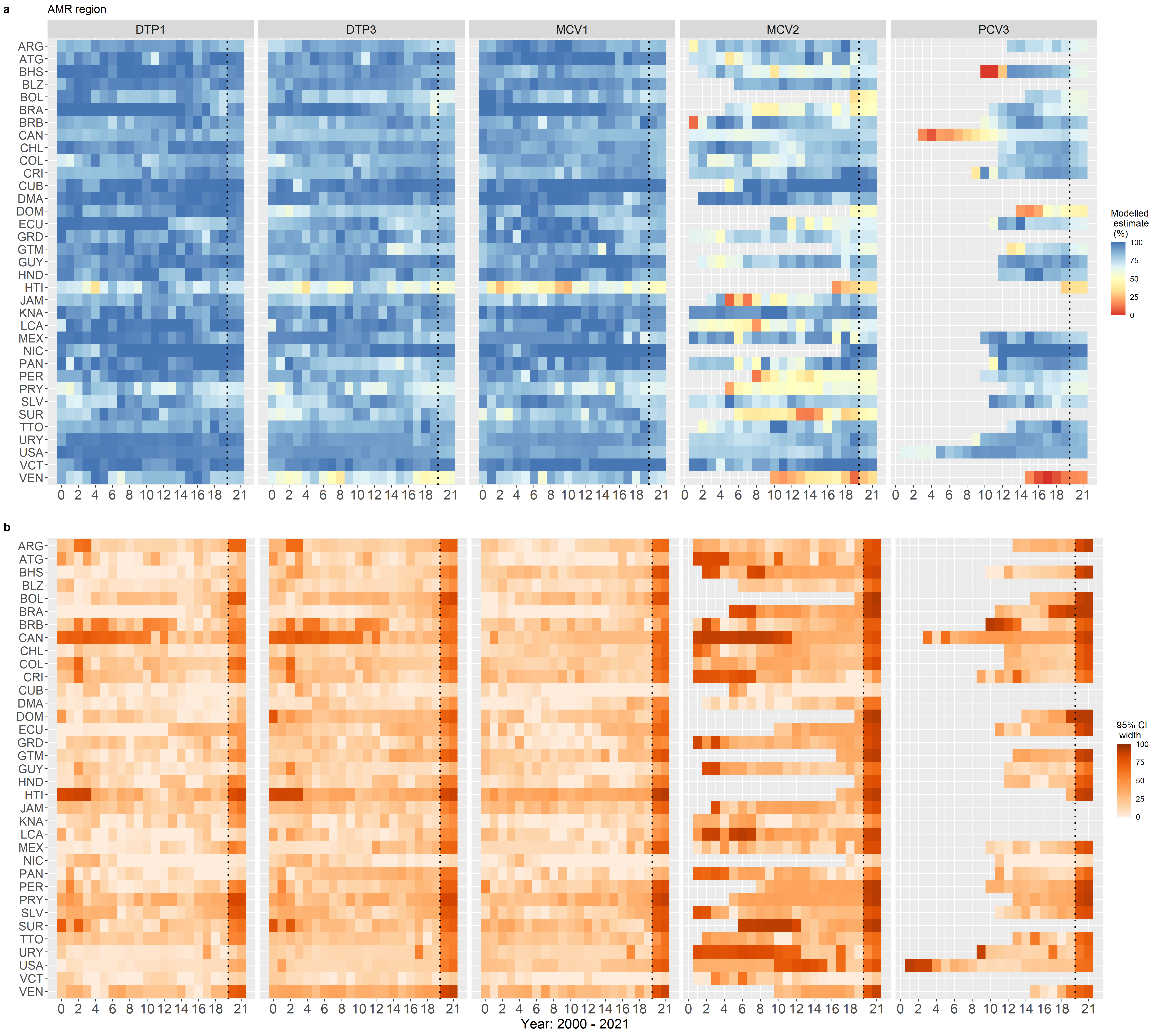}
    \caption{Plots of modelled estimates and corresponding uncertainties at the country level for AMR}
\end{figure}


\begin{figure}[H]
    \includegraphics[width = \textwidth]{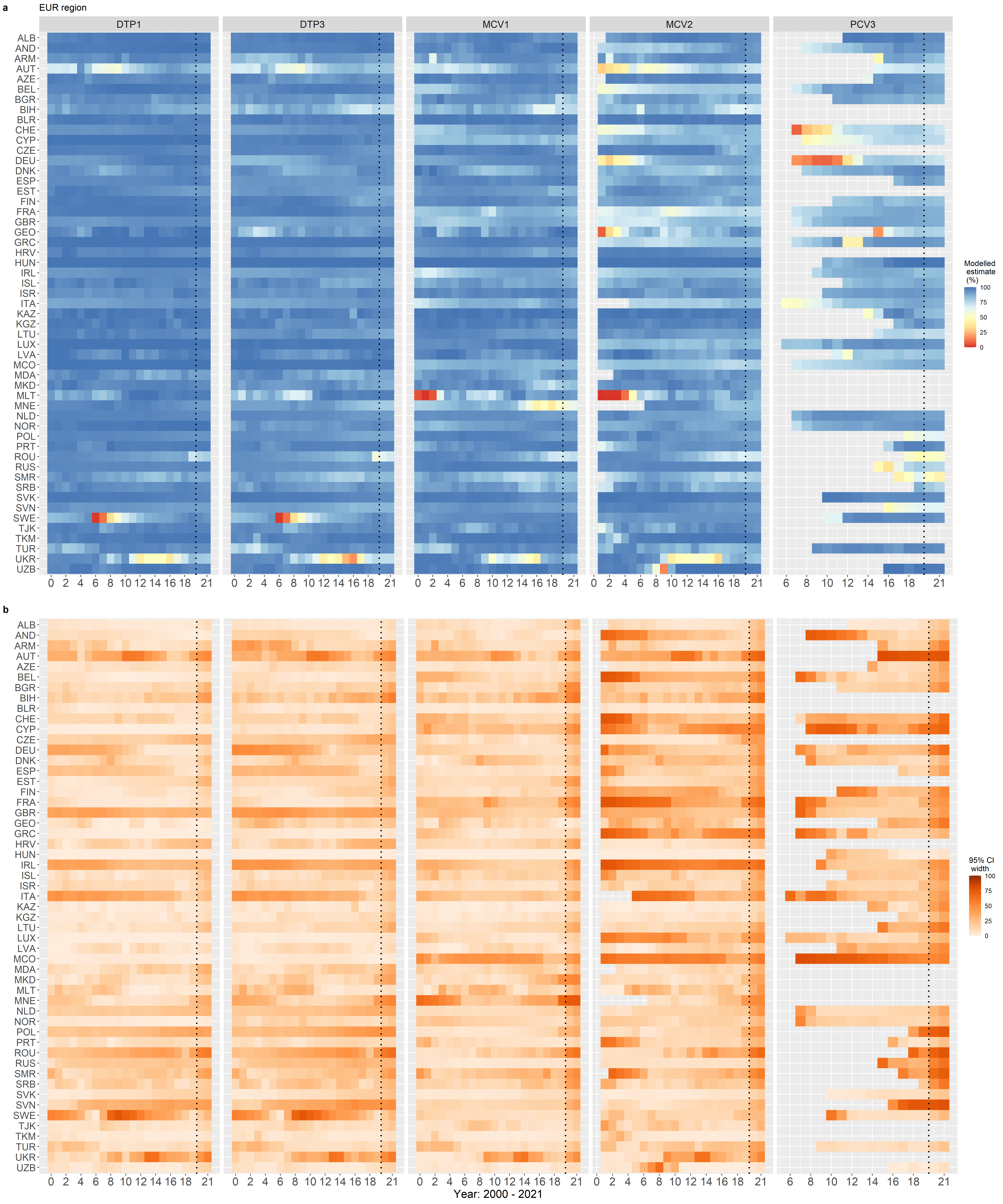}
    \caption{Plots of modelled estimates and corresponding uncertainties at the country level for the EUR}
\end{figure}

\begin{figure}[H]
    \includegraphics[width = \textwidth]{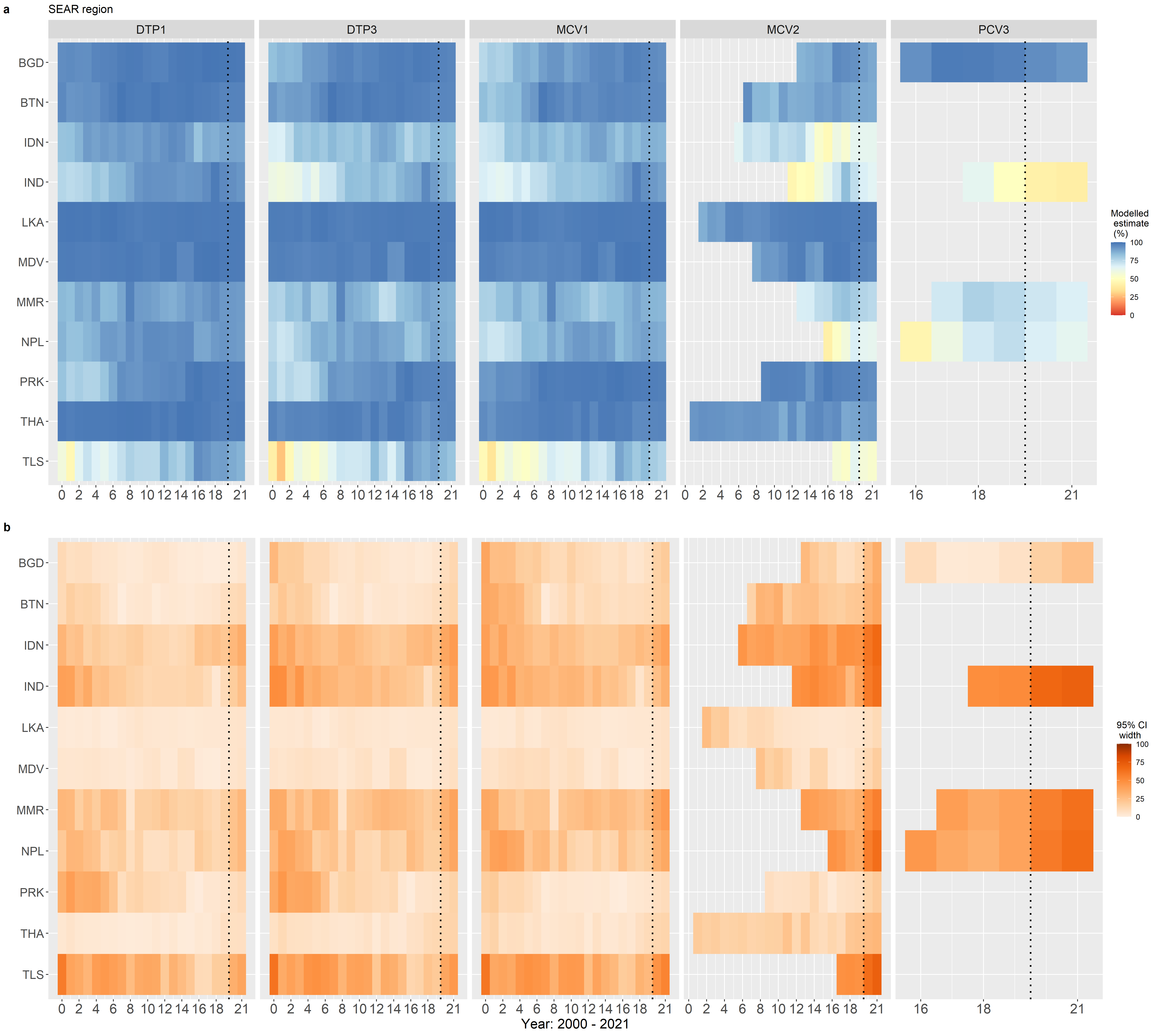}
    \caption{Plots of modelled estimates and corresponding uncertainties at the country level for the SEAR}
\end{figure}

\begin{figure}[H]
    \includegraphics[width = \textwidth]{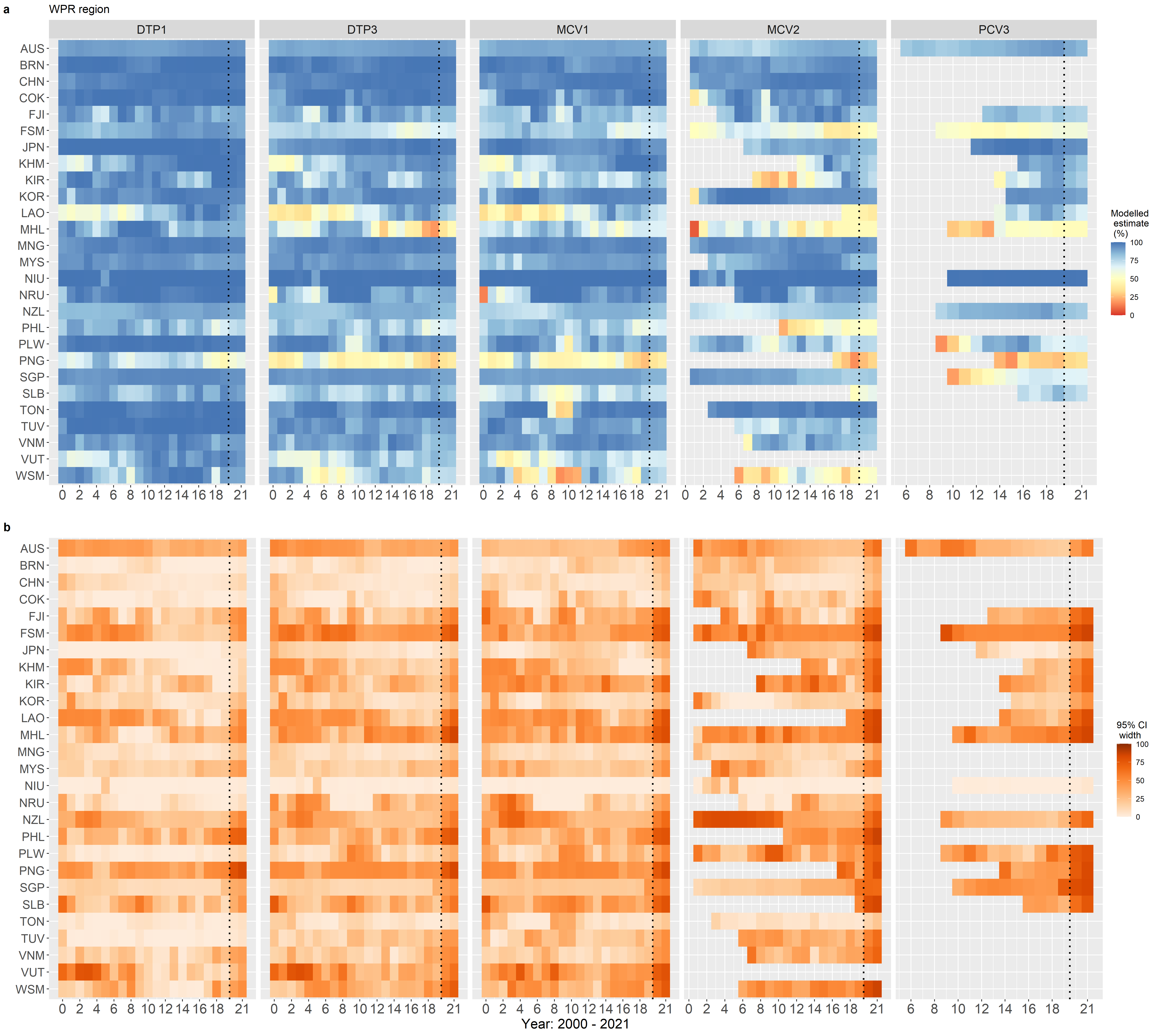}
    \caption{Plots of modelled estimates and corresponding uncertainties at the country level for the WPR}
\end{figure}

\newpage
\begin{figure}[H]
    \centering
 \includegraphics[width=1.0 \textwidth]{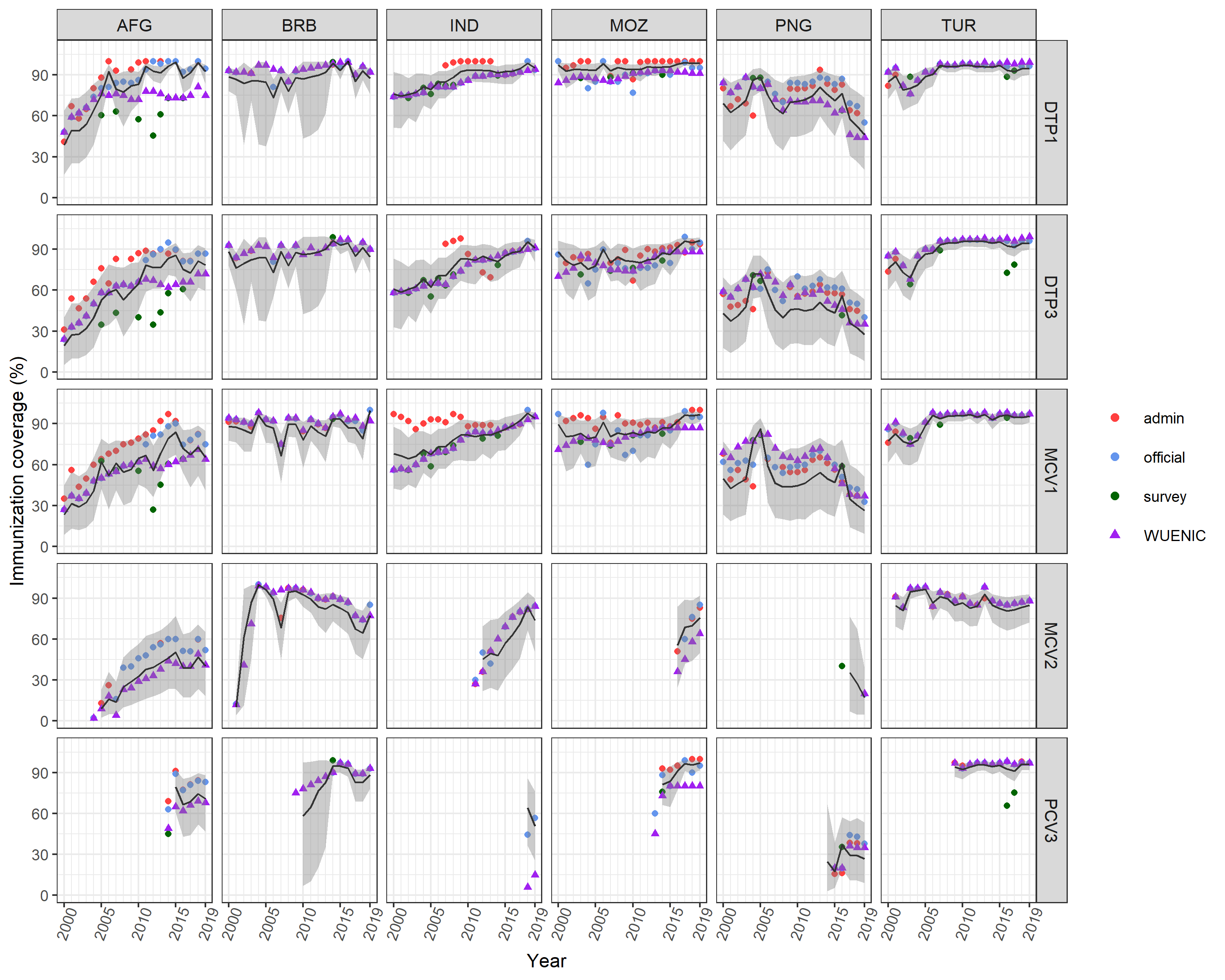} 
 \caption{Modelled estimates of immunization coverage (black line) and 
corresponding uncertainty estimates (grey shaded area) for select countries. The plots are overlaid with processed input administrative, official and survey data, as well as corresponding WUENIC estimates. Predictions for 2021 and 2022 are shown on the right-hand side of the dotted vertical lines.
\label{fig:wueniccomp1}} 
\end{figure}

\clearpage
\newpage
\begin{figure}[H]
\centering
\begin{tabular}{l}
\includegraphics[width=0.9\textwidth]{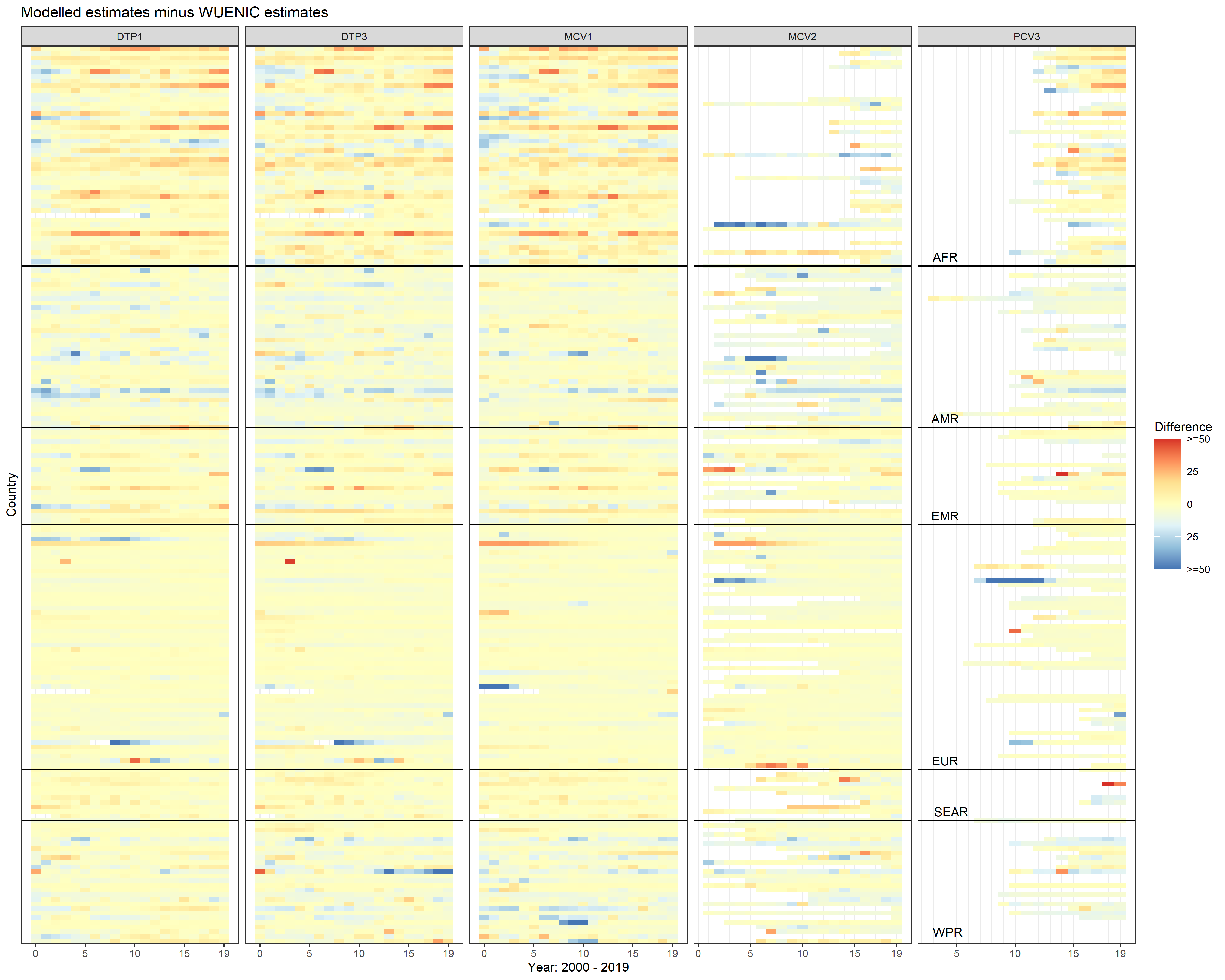}
\end{tabular}
\caption{Differences between WUENIC estimates and the modelled estimates for all WHO countries.} 
\label{Fig:wueniccomp2} 
\end{figure}

\newpage
\begin{figure}[H]
    \includegraphics[width = \textwidth]{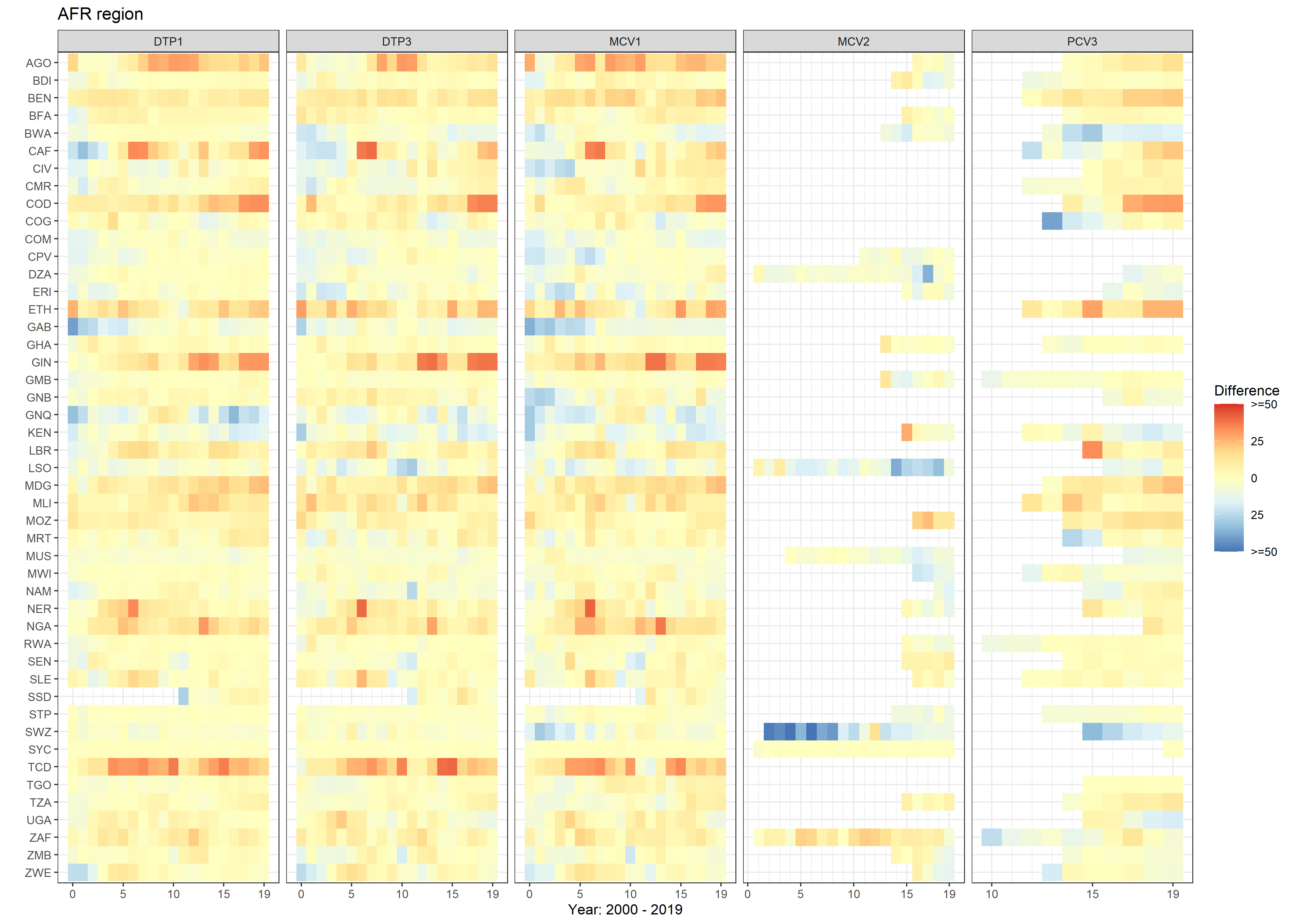}
    \caption{Comparing WUENIC estimates with the modelled estimates for AFR}
\end{figure}

\begin{figure}[H]
    \includegraphics[width = \textwidth]{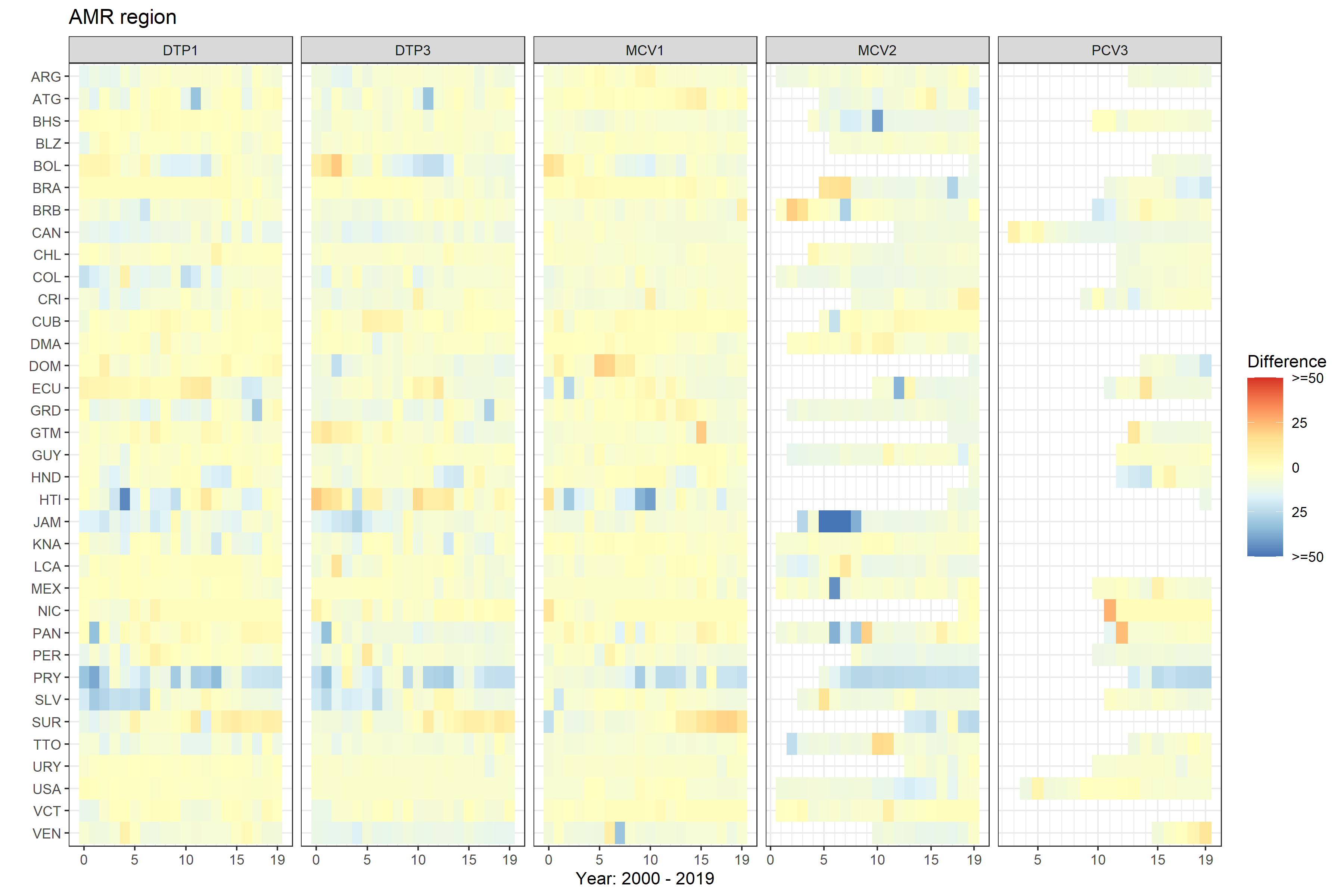}
    \caption{Comparing WUENIC estimates with the modelled estimates for AMR}
\end{figure}

\begin{figure}[H]
    \includegraphics[width = \textwidth]{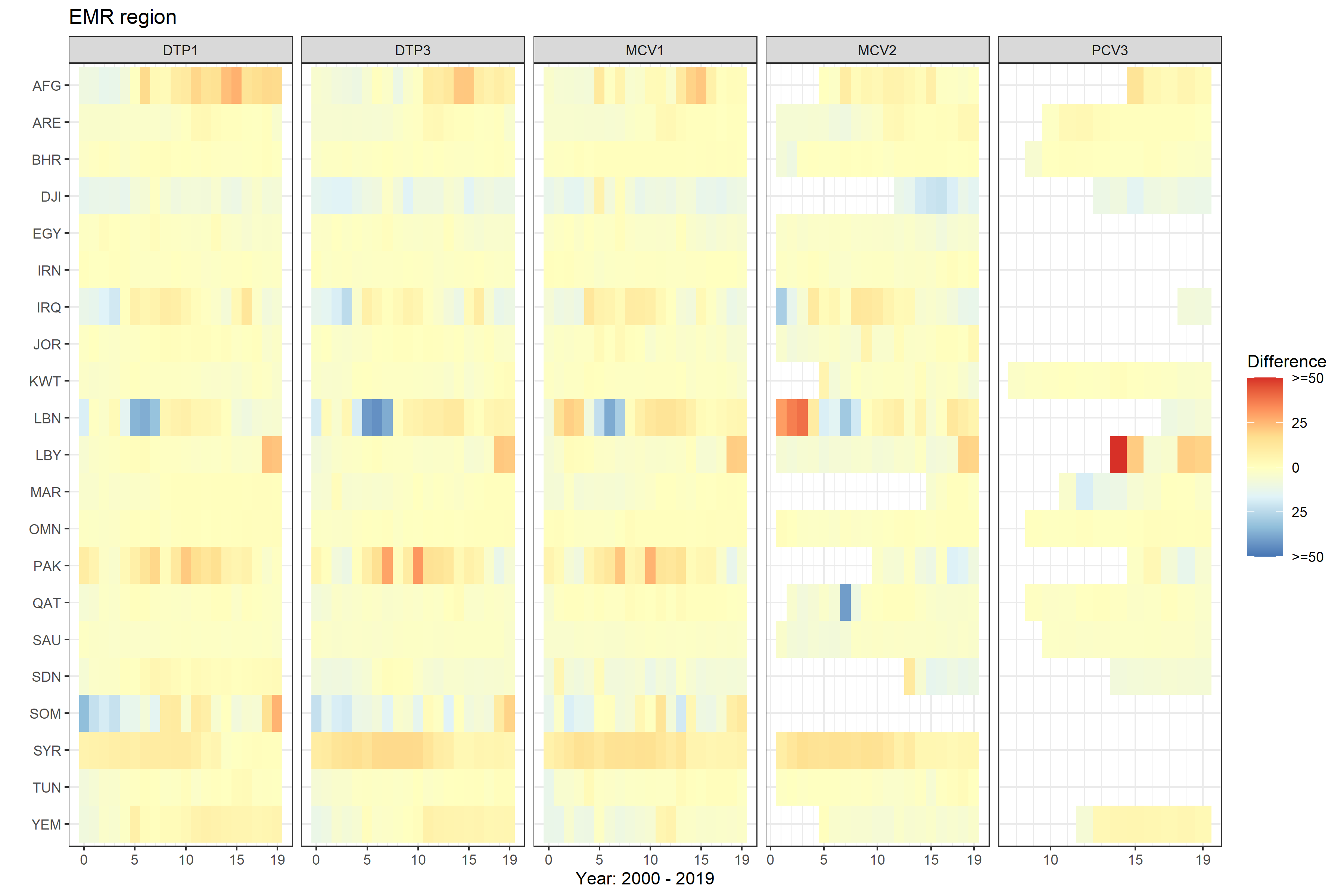}
    \caption{Comparing WUENIC estimates with the modelled estimates for EMR}
\end{figure}

\begin{figure}[H]
    \includegraphics[width = \textwidth]{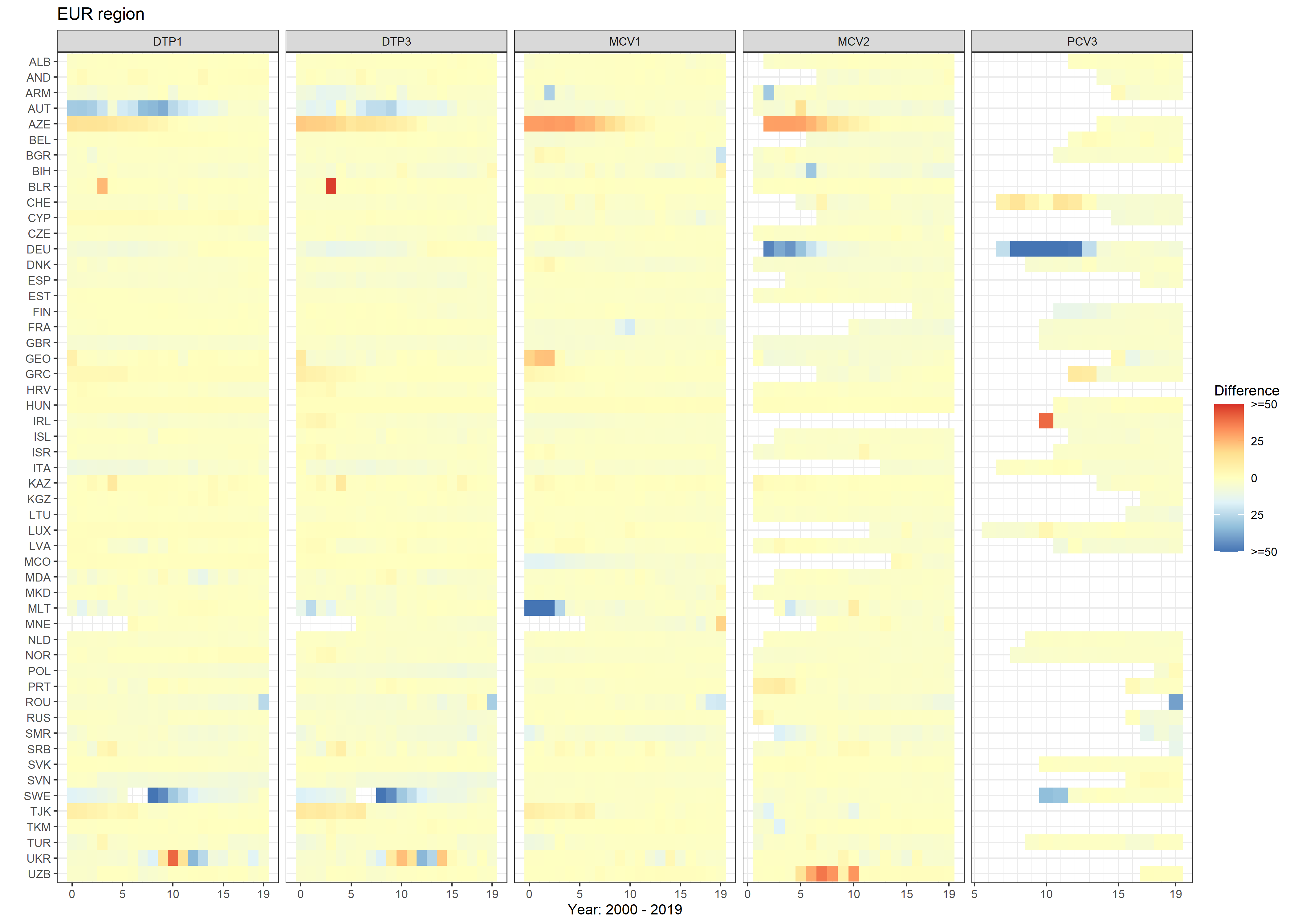}
    \caption{Comparing WUENIC estimates with the modelled estimates for EUR}
\end{figure}

\begin{figure}[H]
    \includegraphics[width = \textwidth]{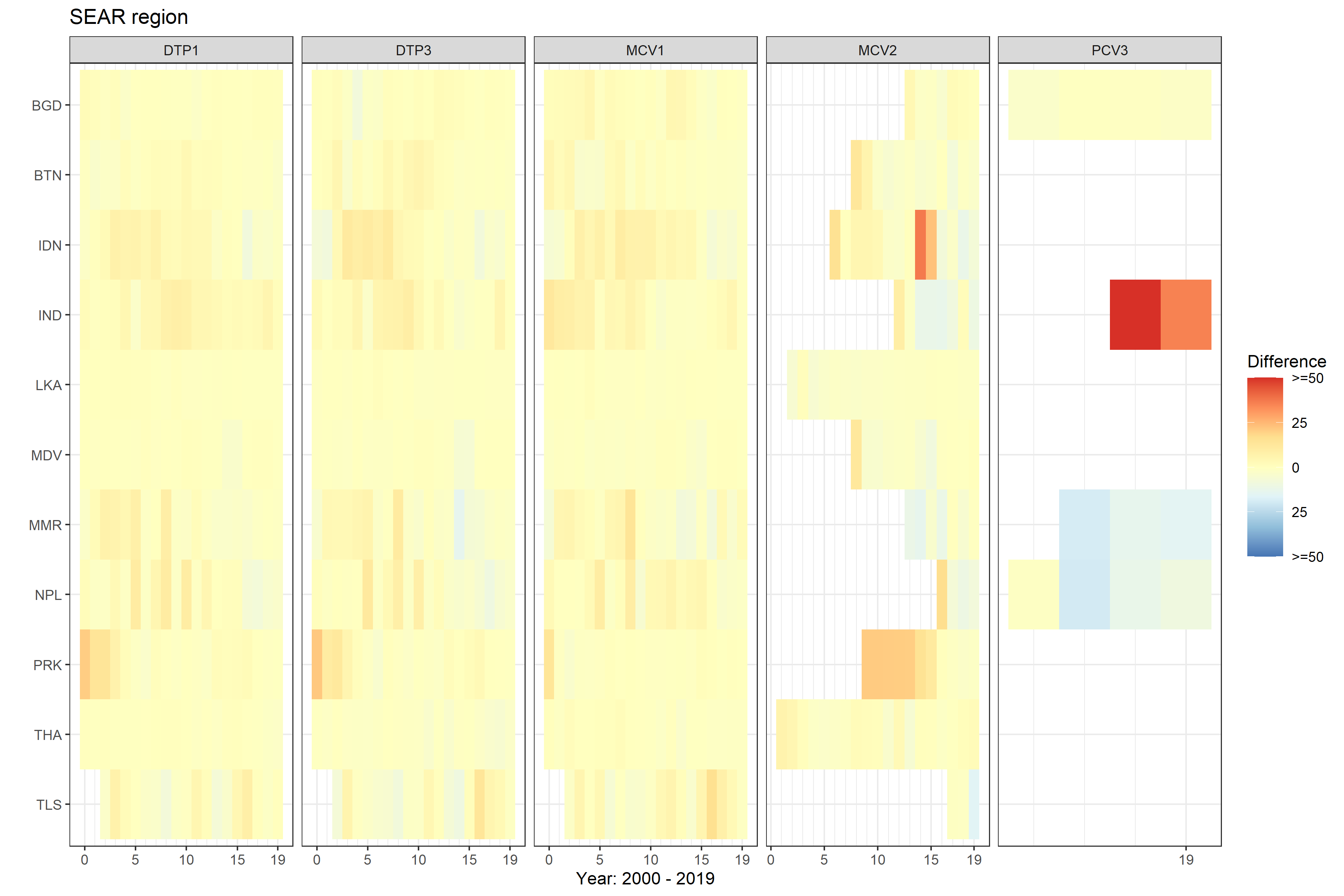}
    \caption{Comparing WUENIC estimates with the modelled estimates for SEAR}
\end{figure}

\begin{figure}[H]
    \includegraphics[width = \textwidth]{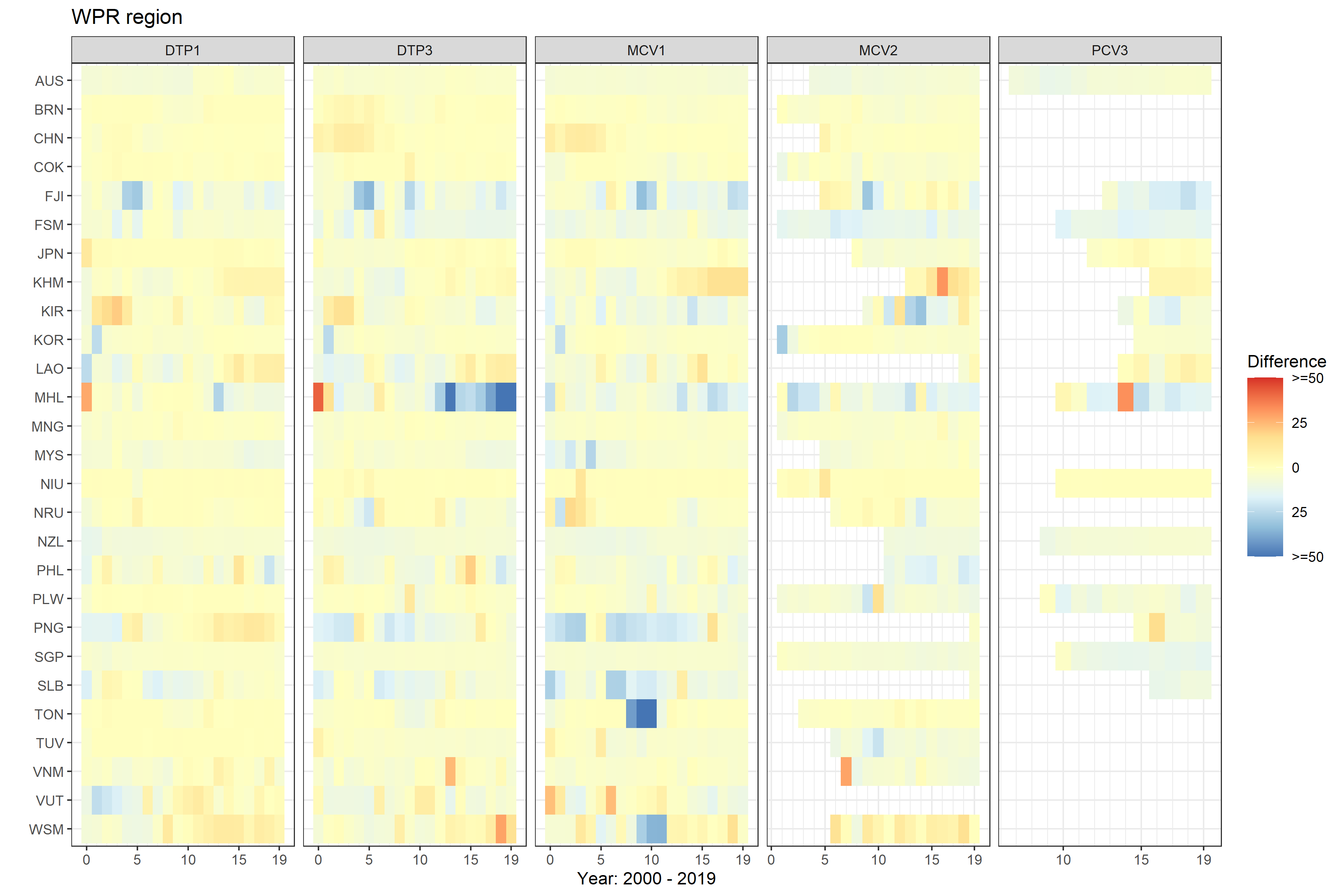}
    \caption{Comparing WUENIC estimates with the modelled estimates for WPR}
\end{figure}

\begin{figure}[htp]
    \centering
 \includegraphics[width=1.0\textwidth]{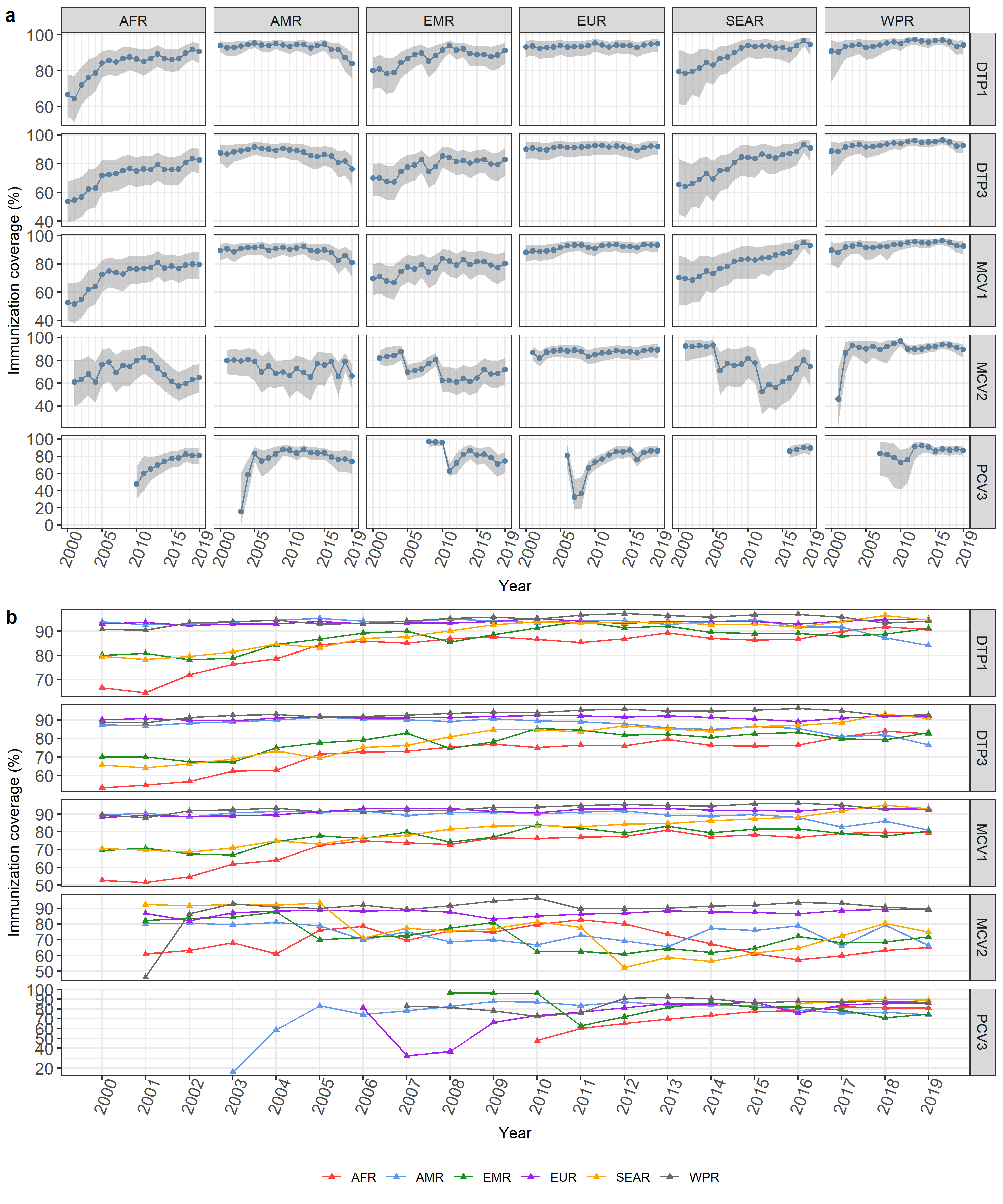} 
 \caption{(a) Modelled estimates of immunization coverage and corresponding uncertainties for each vaccine and WHO region; (b) Regional trends in modelled estimates of coverage for each vaccine.
\label{fig:regfig}} 
\end{figure}

\clearpage
\section{Processing code}
\begin{lstlisting}[language=R]
# Global immunisation coverage modelling

library(imcover)
library(rstan)

options(mc.cores = parallel::detectCores()) 

# path to files - note: change here
dir <- '/main/path/to/workspace'

# get WUENIC input files
cov <- download_coverage(use_cache = TRUE)
svy <- download_survey(use_cache = TRUE)

# drop some categories (PAB, HPV, WUENIC) and clean up
cov <- cov[cov$coverage_category %in% c("ADMIN", "OFFICIAL"), ]
cov$coverage_category <- tolower(cov$coverage_category)

# combined dataset
dat <- rbind(cov, svy)

# remove records with missing values
dat <- dat[!is.na(dat$region), ]
dat <- dat[!is.na(dat$coverage), ]

# mismatch in vaccine names between coverage and survey datasets
dat[dat$antigen == 'DTPCV1', 'antigen'] <- 'DTP1'
dat[dat$antigen == 'DTPCV2', 'antigen'] <- 'DTP2'
dat[dat$antigen == 'DTPCV3', 'antigen'] <- 'DTP3'

# select only some vaccines to analyse
dat <- ic_filter(dat, vaccine = c("DTP1", "DTP3", "MCV1", "MCV2", "PCV3"))
  table(dat$coverage_category, dat$antigen)

# subset the time
dat <- ic_filter(dat, time = 2000:2020)

# drop observations with zero values
  table(dat$coverage == 0)  # n=165
dat <- dat[dat$coverage > 0, ]

# adjustment - use ratio for DTP3 and requiring 0 < ic < 100
dat <- ic_ratio(dat, numerator = 'DTP3', denominator = 'DTP1')
dat <- ic_adjust(dat, coverage_adj = TRUE)

# model fitting - multiple likelihood model
fit <- ic_fit(dat,
              chains = 4,
              iter = 2000,
              warmup = 1000,
              prior_lambda = c(0.5, 0.5, 0.5),
              prior_sigma = c(2, 2, 0.2),
              upper_sigma = c(100, 100, 0.4),
              lower_sigma = c(0, 0, 0))

# save the fitted object
saveRDS(fit, file = file.path(dir, 'output', 'fit.rds'))

# post-processing
regs <- names(fit) # regions

# WAIC calculation
waics <- lapply(regs, function(r){
  print(r)
  ww <- waic(fit[[r]])
  print(ww)
})

# coverage estimates and predictions
# loop over the regions in the fitted object
for(r in regs){
  print(r)
  # extract region
  f <- fit[[r]]
  
  # filter based on year of introduction
  f <- filter_yovi(f)
  
  # predict for 2021 and 2022
  f <- predict(f, t=2)
  
  # summarise coverage estimates
  ic_post <- ic_coverage(f, 
                         object = 'posterior',
                         stat = c('mean', 'quantile'),
                         probs = c(0.025, 0.5, 0.975))
  # and predictions
  ic_pred <- ic_coverage(f,
                         object = 'prediction',
                         stat = c('mean', 'quantile'),
                         probs = c(0.025, 0.5, 0.975))
  
  ic <- rbind(ic_post, ic_pred)
  ic <- ic[order(ic$country, ic$vaccine, ic$time),]
  
  # write out table of coverage
  write.csv(ic, file.path(dir, 'output', paste0(r, '_coverage.csv')),
            row.names = F)
}

# regional aggregation summaries
# first load the population denominator
denom <- read.csv(file.path(dir, 'input', 'denom.csv'))
# this was created with: denom <- imcover::download_wuenic()

# process each region ...
for(r in regs){
  print(r)
  
  # population-weighted summary
  icr <- ic_regional(fit[[r]],
					 denom = denom,
					 filter_yovi = F)

  # create output table					 
  write.csv(icr,
			file.path(dir, 'output', paste0(r, '_regional_coverage.csv')),
			row.names = F)
}
\end{lstlisting}
\bibliographystyle{rss} 
\bibliography{REFERENCES}